\documentclass[a4paper,fleqn,usenatbib]{mnras}

\usepackage[T1]{fontenc}
\usepackage{ae,aecompl}
\usepackage{adjustbox}

\usepackage{graphicx}	
\usepackage{amsmath}	
\usepackage{amssymb}	
\usepackage{lscape}
\usepackage{pdflscape}
\usepackage[flushleft]{threeparttable}
\usepackage{multirow}

\title[Spectral Features of Red Clumps]{Identification, mass and age of primary red clump stars from spectral features derived with the LAMOST DR7}

\author[X.-J. He et al.]{
Xu-Jiang He,$^{1,2}$
A-Li Luo,$^{1,2}$\thanks{E-mail: lal@nao.cas.cn}
Yu-Qin Chen$^{1,2}$
\\
$^{1}$CAS Key Laboratory of Optical Astronomy, National Astronomical Observatories, Chinese Academy of Sciences, Beijing 100101, China\\
$^{2}$School of Astronomy and Space Science, University of Chinese Academy of Sciences, Beijing 100049, China\\
}
\date{Accepted 2022 February 17. Received 2022 February 17; in original form 2021 November 9}
\pubyear{2021}

\begin{document}
\label{firstpage}
\pagerange{\pageref{firstpage}--\pageref{lastpage}}
\maketitle

\begin{abstract}
Although red clump (RC) stars are easy to identify due to their stability of luminosity and color, about 20-50\% are actually red giant branch (RGB) stars in the same location on the HR diagram. In this paper, a sample of 210 504 spectra for 184 318 primary RC (PRC) stars from the LAMOST DR7 is identified, which has a purity of higher than 90 percent. The RC and the RGB stars are successfully distinguished through LAMOST spectra(R$\sim$1800 and SNR$>$10) by adopting the XGBoost ensemble learning algorithm, and the secondary RC stars are also removed. The SHapley Additive exPlanations (SHAP) value is used to explain the top features that the XGBoost model selected. The features are around Fe5270, MgH \& Mg Ib, Fe4957, Fe4207, Cr5208, and CN, which can successfully distinguish RGB and RC stars. The XGBoost is also used to estimate the ages and masses of PRC stars by training their spectra with Kepler labeled asteroseismic parameters. The uncertainties of mass and age are 13 and 31 percent, respectively. Verifying the feature attribution model, we find the age-sensitive elements XGBoost gets are consistent with the literature. Distances of the PRC stars are derived by $K_{S}$ absolute magnitude calibrated by Gaia EDR3, which has an uncertainty of about 6 percent and shows the stars mainly locate at the Galactic disk. We also test the XGBoost with R$\sim$250, which is the resolution of the Chinese Space Station Telescope(CSST) under construction, it is still capable of finding sensitive features to distinguish RC and RGB.

\end{abstract}




\section{Introduction}
\label{sec:info}

Red clump (RC) stars are low-mass stars in the stage of core helium-burning \citep{cassisi1997critical}. They are close to the first-ascent red giant branch (RGB) and have sizable convective envelopes resulting from a significant buffer of mass (a few 0.1M$_\odot$) above the hydrogen-burning shell \citep{girardi2016}. Their luminosity and color are very stable, which can be used as a standard candle to measure distance \citep{paczynski1998,girardi1998,alves2000,groenewegen2008}. At the same time, they are widely distributed, with high brightness and easy to observe. These make RC become an important probe to explore the structure and evolution of the Galaxy \citep{lopez2002}. There have been many previous efforts to identify RC in large spectral and asteroseismic survey data, such as the Gaia space mission \citep{gaia2016,gaia2018}, the LAMOST Experiment for Galactic Understanding and Exploration (LEGUE) \citep{deng2012,zhao2012}, the Apache Point Observatory Galactic Evolution Experiment (APOGEE) \citep{majewski2017apache}, the GALAH \citep{de2015galah} and Kepler mission.

Due to the stability of luminosity and color, the RC stars form a high density distribution area in the $T \rm _{eff}$-log $g$ diagram, which is easy to identify \citep{lopez2002}. However, there will be 20 - 50\% pollution which mainly comes from the red giant branch (RGB) stars in the same location. RGB stars burn hydrogen in the shell around the inert helium core, while RC stars burn helium core \citep{girardi1998}. We can use the asteroseismic parameters (the large frequency separation between p-modes, $\triangle \nu$; the period spacing of the mixed g- and p-modes, $\triangle P$) to separate the two kinds of stars clearly \citep{bedding2011}. To get asteroseismic parameters one needs many observations over a long period of time. Kepler first phase of the sky survey is to observe the 105 square degrees sky area between Cygnus and Lyra for four years. It provides high-precision photometric data, which can be used to calculate the stellar seismological parameters -- $\triangle \nu$ and $\triangle P$ \citep{stello2013,mosser2014,vrard2016,elsworth2017}. However, this method is only suitable for a small number of giant stars whose asteroseismic parameters can be measured, and can not obtain large scale RC stars. The two kinds of stars can also be distinguished by spectral classification method, mainly according to the comparison of $T \rm _{eff}$ – log $g$ – [Fe/H] and their color–magnitude compared to theoretical expectation from isochrones, the stars can be classified as RC or RGB \citep{bovy2014}. \cite{huang2020mapping} have successfully selected 140,000 primary red clump (RC) stars with purity and completeness generally higher than 80 per cent from LAMOST DR4. This method needs the accuracy of $T \rm _{eff}$ and log $g$ are better than 100K and 0.1dex respectively, which is a difficult condition to meet, especially for low-resolution spectroscopy. Some studies also pointed out that because of the non-canonical extra mixing processes, the C and N abundances in the photosphere will change continuously when RGB stars evolve to RC stars \citep{charbonnel1995,charbonnel1997,martell2008,masseron2016,hawkins2018}. The abundances of elements will be reflected in the spectra, so if we can distinguish RC stars by spectra, we will get a larger number of RC samples with higher purity \citep{hawkins2018,ting2018,wu2019}. 

In order to understand the stellar population and structure of the Milky Way through RC stars, multi-dimensional information of star samples is needed, such as 3D positions, mass, age, and elemental abundances. Among these stellar parameters, age and distance are important but hard to acquire directly. The age of a star can be obtained by matching its atmospheric parameters with stellar isochrones, but this method is only applicable to main-sequence turn-off and sub-giant stars \citep{xiang2017,mints2017,yu2018}. This method is not suitable for RC stars, because the distribution of stellar isochrones in the region where RC appears in the HR diagram is too dense to give accurate ages. For low-mass giant stars, we can obtain a good age estimate by the accurate mass of the star \citep{ness2016,martig2016,wu2018}. Masses and ages of stars can also be reflected in the spectra \citep{martig2016,ness2016,sanders2018}. Due to the first-dredge-up process, the carbon to nitrogen abundance ratio [C/N] is tightly correlated with stellar mass. This means [C/N] ratios deducible from the spectra, and thus can be further used to derive their ages. RC stars have intrinsic absolute magnitude, so they can be used as standard candles for distance measurement. In fact, according to the stellar evolution model, RC stars do not always keep the same luminosity. RC stars are divided into two different stellar populations: primary RC (PRC) stars, which have electronic degenerate cores, and the second RC (SRC) stars, which contain non-degenerate He-cores \citep{girardi1999}.The difference between them is that the SRC stars are massive and younger than 1Gyr. The most important thing is that SRC stars do not have intrinsic absolute magnitude, so they are not suitable for distance estimation.

In this paper, we aim to identify high purity primary RC stars from the LAMOST DR7 low-resolution spectrum and then extract and analyze important spectral features related to RGB and RC stars. Spectral feature extraction is a process of decomposing, recombining, and selecting spectral data components, which is a key link in spectral data mining. It not only determines the quality, efficiency, system complexity, and robustness of subsequent data processing but also relates to what knowledge can be mined and the interpretability of the physical meaning of processing results. Finally, the mass, age, and distance are also provided for the sample stars where the mass and age are determined from the LAMOST low resolution spectra, and the distances are derived from the $K_S$ absolute magnitudes.

The paper is organized as follows. In Section 2, we introduce the data and machine learning methods used in the current analysis. Then, we introduce the selection of PRC stars in Section 3 and analyze the spectral features related to classification. In Section 4, we calculate the mass and age of the PRC sample. In Section 5, the distances of the PRC sample are calculated. Statistical properties of the PRC samples are presented in Section 6. Finally, discussion and summary are given in Section 7.

\section{Data and Method}
\label{sec:data} 
\subsection{Data}
The LAMOST telescope \citep{cui2012large} has released the seventh data release (DR7) in 2021, including more than 10 million stellar spectra of its Galactic Survey \citep{deng2012,zhao2012,liu2015preface}. In this work, the labels of effective temperature $T \rm _{eff}$, surface gravity log $g$ and metallicity [Fe/H] we used are directly from the LAMOST released catalog which is derived by the LAMOST Stellar Parameter pipeline (LASP)\citep{wu2011automatic,luo2015first}. We select the stars from LAMOST DR7 with specific interval of log $g$, $T \rm _{eff}$ and [Fe/H], to widely cover the RC region..  Finally, we obtain 819,657 stars with signal-to-noise ratio (SNR) higher than 10.

The wavelength range of LAMOST low-resolution spectra are from 3690 to 9100\AA\, and the resolution power is $R \sim 1800$ at the wavelength of 5500\AA. Each spectrum is rebinned with a 1\AA~step onto the wavelength range from 3800\AA\ to 9000\AA~using a cubic non-linear interpolator, and is scaled to eliminate scale differences among the raw spectra as below. 

\begin{equation}
    \tilde{x^{i}} = \frac{x^{i}-\mu}{\sigma},
	\label{eq:t1}
\end{equation}

where $x^{i}$ represents a spectrum, $\mu$ indicates the mean value of the $x^{i}$, $\sigma$ indicate the standard deviation of the $x^{i}$, and the scaled spectrum is denoted as $\tilde{x}^{i}$.

In addition, we use Kepler data to construct three asteroseismic samples for training data labeling, stellar mass estimation and result verification. We use the asteroseismic samples provided by \cite{vrard2016} to mark the evolutionary stages of stars in the training set, including PRC, SRC and RGB. At the same time, we used the asteroseismic samples provided by \cite{elsworth2017} to verify the results of the classification model prediction. Finally, we use the asteroseismic parameters $\nu_{\rm max}$ and $\triangle \nu$ measured by \cite{yu2018} to estimate the mass. Based on the end-of-mission long-cadence data of Kepler, \cite{yu2018} systematically characterized solar-like oscillations and granulation of 16,094 oscillating red giants, and the obtained parameters were very accurate.

\subsection{Method}
\label{sec:method}
In the past decades, many machine learning methods (such as ANNs, SVM, KNN, KPCA, etc.) have been successfully applied to stellar spectral classification and physical parameters estimation. XGBoost proposed by \cite{chen2016xgboost} is a supervised learning algorithm realized by gradient tree boosting, which can solve machine learning problems such as classification and regression. XGBoost optimizes the gradient boosting framework \citep{friedman2001greedy,friedman2002stochastic} and has a faster parallel processing speed and higher accuracy than the traditional decision tree. Since it was proposed in 2016, it has been widely used in various fields such as price prediction \citep{gumus2017crude,jabeur2021forecasting}, disease diagnosis \citep{ogunleye2019xgboost,pang2019novel}, fault detection \citep{zhang2018data,chakraborty2019early} and so on because of its high accuracy and no less than that of ANNs.

\begin{figure}
	\centering
	\includegraphics[width=\columnwidth]{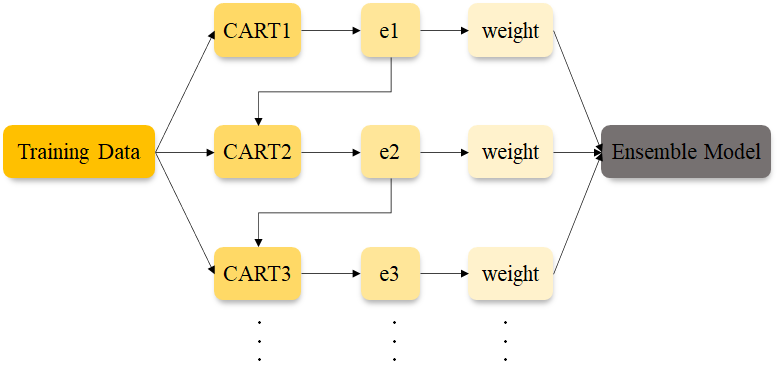}
    \caption{Boosting iterative process of XGBoost.}
    \label{fig:fig1}
\end{figure}

XGBoost is composed of multiple CARTs (Classification And Regression Tree). Each decision tree learns the residual of the sum of the target value and the predicted value of all previous trees. Multiple decision trees make decisions together, and finally add up the results of all trees as the final prediction result as shown in Fig. \ref{fig:fig1}. In the training stage, each new tree is trained on the basis of the previously trained tree. Each tree is generated by the idea of binary recursive splitting: firstly, the loss function is set to traverse the feature and the segmentation point, and the sample is divided into two sub sample sets by the feature and the segmentation point with the most reduced loss function, and then the process is recursive to the sub set until the end condition is satisfied. The prediction value of the model is stored on the leaf node of the tree, and the average value of most samples in the leaf node is usually taken as the prediction value of the node. In the testing stage, the samples to be identified are put into the model, starting from the root node of each tree, passing through the branches to a leaf according to the conditions, and then the corresponding results are obtained.

Compared with a linear model, XGBoost has better accuracy in prediction, but it also loses the interpretability of the linear model. For complex models (such as integration models or deep learning models), we can’t use the original model itself as its interpretation. On the contrary, we need to use a simple interpretation model, which is defined as any interpretation approximation of the original model. In order to explain the tree model, there are some calculation methods of feature importance, such as gain, split, cover and SHAP \citep{lundberg2017unified,lundberg2017consistent}. In this work, we measure the importance of the features by the SHAP value. Compared with other calculation methods, the advantage of SHAP is that it has better consistency, and it can analyze each specific sample, not only give global special importance. Different samples have different SHAP values for the same feature. The final output value of the sample in the decision tree can be expressed as the sum of the SHAP values of each feature of the sample (plus the basic score). 

The prediction value of the model is interpreted as the sum of the importance of each input feature. The relation between the predicted value and importance scores is performed as
\begin{equation}
    f(x) = g(x) = \phi_{0} + \sum_{i=1}^{M}{\phi_{i}},
	\label{eq:t2}
\end{equation}
where $f$ is the original prediction model to be explained and $g$ is the explanation model, $f(x)$ stands for the predicted value of the input sample $x$ in the decision tree. Here $M$ is the number of features, $\phi_{i}$ is the attribution of each feature, $\phi_{0}$ is the constant of the interpretation model, that is, the predicted mean value of all training samples. For a specific sample, the calculation of the SHAP value $\phi_i$ of characteristic $i$ is performed as
\begin{equation}
    \phi_i = \sum_{S\subseteq \{x_1,...,x_M\} \backslash \{x_i\}}\frac{|S|!(M-|S|-1)!}{M!}(f_x(S\cup\{x_i\})-f_x(S)),
	\label{eq:t3}
\end{equation}
where $\{x_1,...,x_M\}$ is a collection of all input features, $M$ is the number of features, $\{x_1,...,x_M\} \backslash \{x_i\}$ is a possible set of all input features excluding $\{x_i\}$. $f_x(S)$ is the prediction for feature subset $S$, $f_x(S\cup\{x_i\})$ is the prediction for feature subset $S$ including $\{x_i\}$. $\frac{|S|!(M-|S|-1)!}{M!}$ represents the weight, please refer to \cite{lundberg2017unified} for specific explanation and details. All in all, the meaning of the SHAP value $\phi_i$ is the weighted average of the changes of the predicted values of the model before and after adding feature $\{x_i\}$ for any feature subset without feature $\{x_i\}$. 

\section{Identification of primary RC stars}
\label{sec:identification}

\subsection{Classification of RC and RGB}
\label{sec:classification}

The RC and RGB are two kinds of stars in different evolution stages, but in the $T \rm _{eff}-$log $g$ diagram, they overlap significantly. It's hard to simply distinguish the two kinds of stars by $T \rm _{eff}$ and log $g$. RGB stars burn hydrogen in the shell around the inert helium core, while RC stars burn helium core \citep{girardi1998}. Asteroseismology can diagnose core helium burning in a star, and hence become the most effective standard for distinguishing RC and RGB stars \citep{montalban2010seismic,bedding2011,mosser2011mixed,mosser2012probing,stello2013,pinsonneault2014apokasc,vrard2016,elsworth2017}. 
It is possible to accurately distinguish between RGB stars and RC stars through large frequency intervals ($\triangle \nu$) and periodic intervals ($\triangle P$). However, currently it is difficult to classify the large samples of giant stars by the method of asteroseismology. As introduced in Section 1, the evolution stage of stars can also be reflected in the spectrum by the change of element abundance, so we can get large samples of RC stars by data-driven method.
 
\begin{figure}
    \includegraphics[width=1\columnwidth]{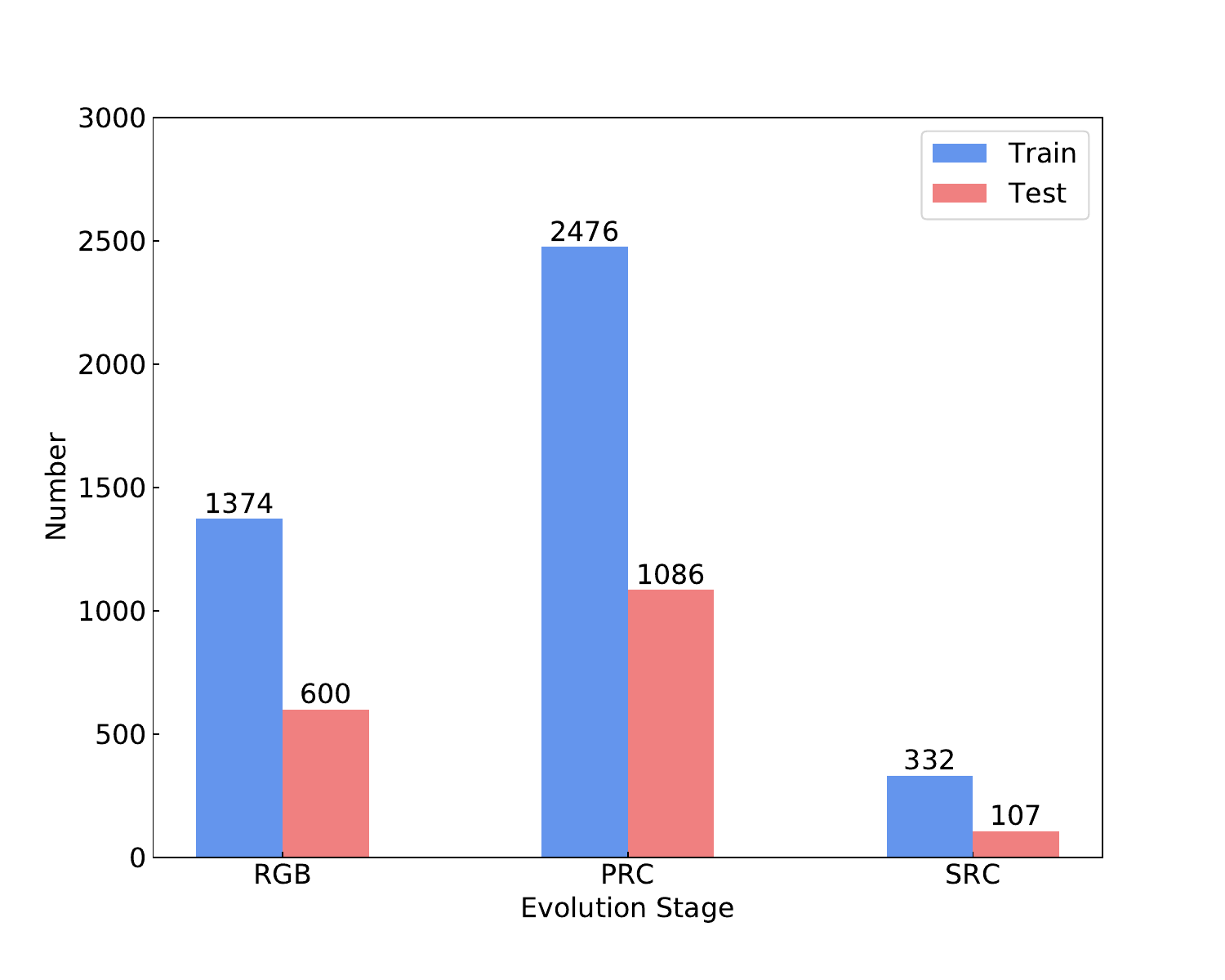}
    \caption{Evolution stage distribution of the training data and test data including RGB, PRC and SRC stars.}
    \label{fig:fig2}
\end{figure}

The exact large frequency and periodic intervals of thousands of stars have been derived from their power spectra by Kepler photometry\citep{stello2013,mosser2014,vrard2016,elsworth2017}.  In this work, we construct a giant stars sample with evolution stages provided by \cite{vrard2016} which was deduced using the asteroseismic information. We cross-match the sample of \cite{vrard2016}  with LAMOST-Kepler stars. A total of 5975 common stars with SNRs higher than 70 with 1974 RGB stars, 3562 PRC stars, and 439 SRC stars are found. In order to ensure that important spectral features are not missed, the range of spectra we used is from 3800 \AA to 9000 \AA, then they were interpolated to an interval of 1 \AA, with total dimensions of 5200. Samples RGB, PRC and SRC were labeled with ``0", ``1", and ``2" respectively. Finally, we randomly select 70\% of the samples as the training data set, and the rest as the test data set as shown in Fig. \ref{fig:fig2}.

\begin{figure}
    \includegraphics[width=1\columnwidth]{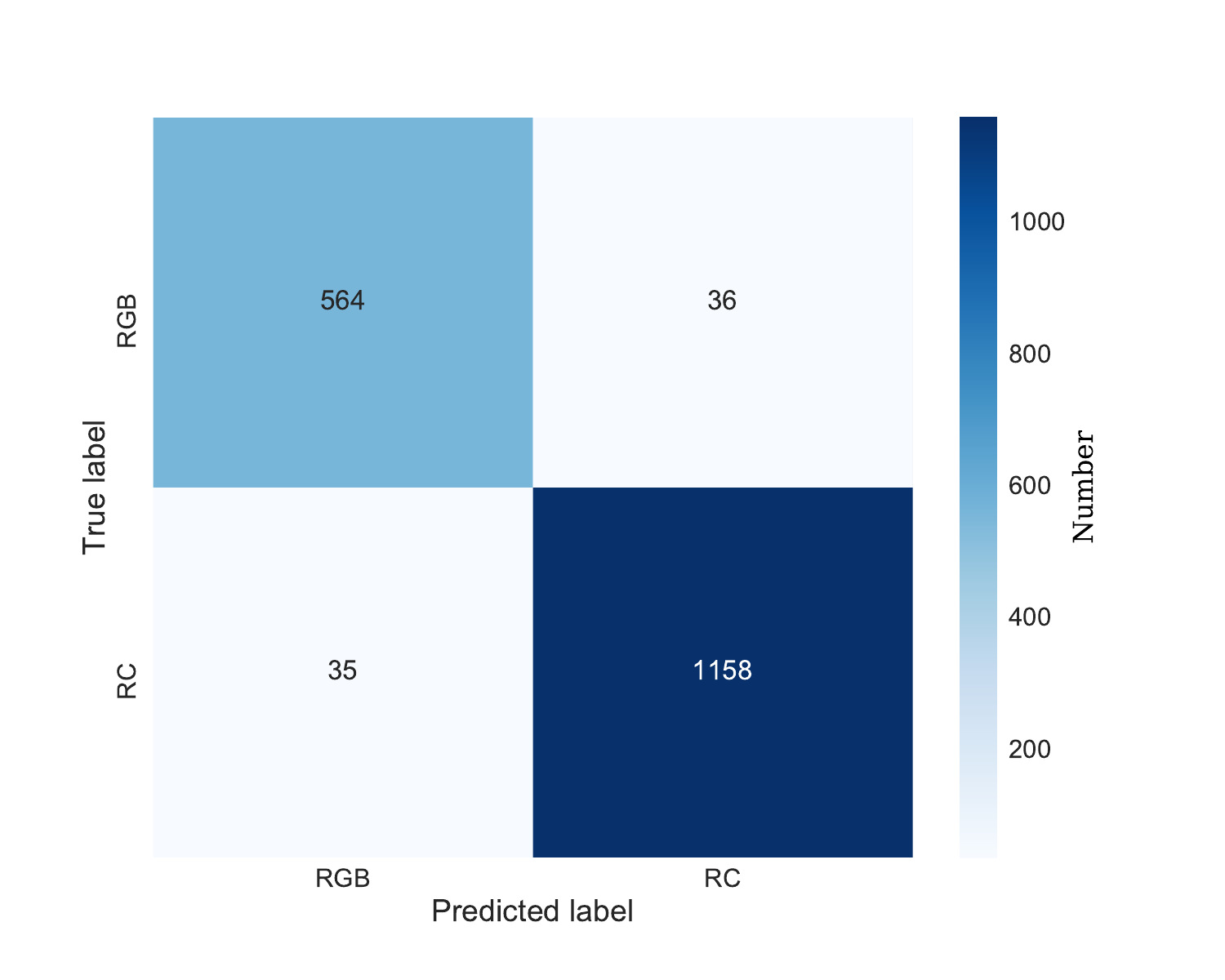}
    \caption{The confusion matrix of the classification model predicted on the test set.}
    \label{fig:matrix}
\end{figure}


\begin{figure*}
    \includegraphics[width=2\columnwidth]{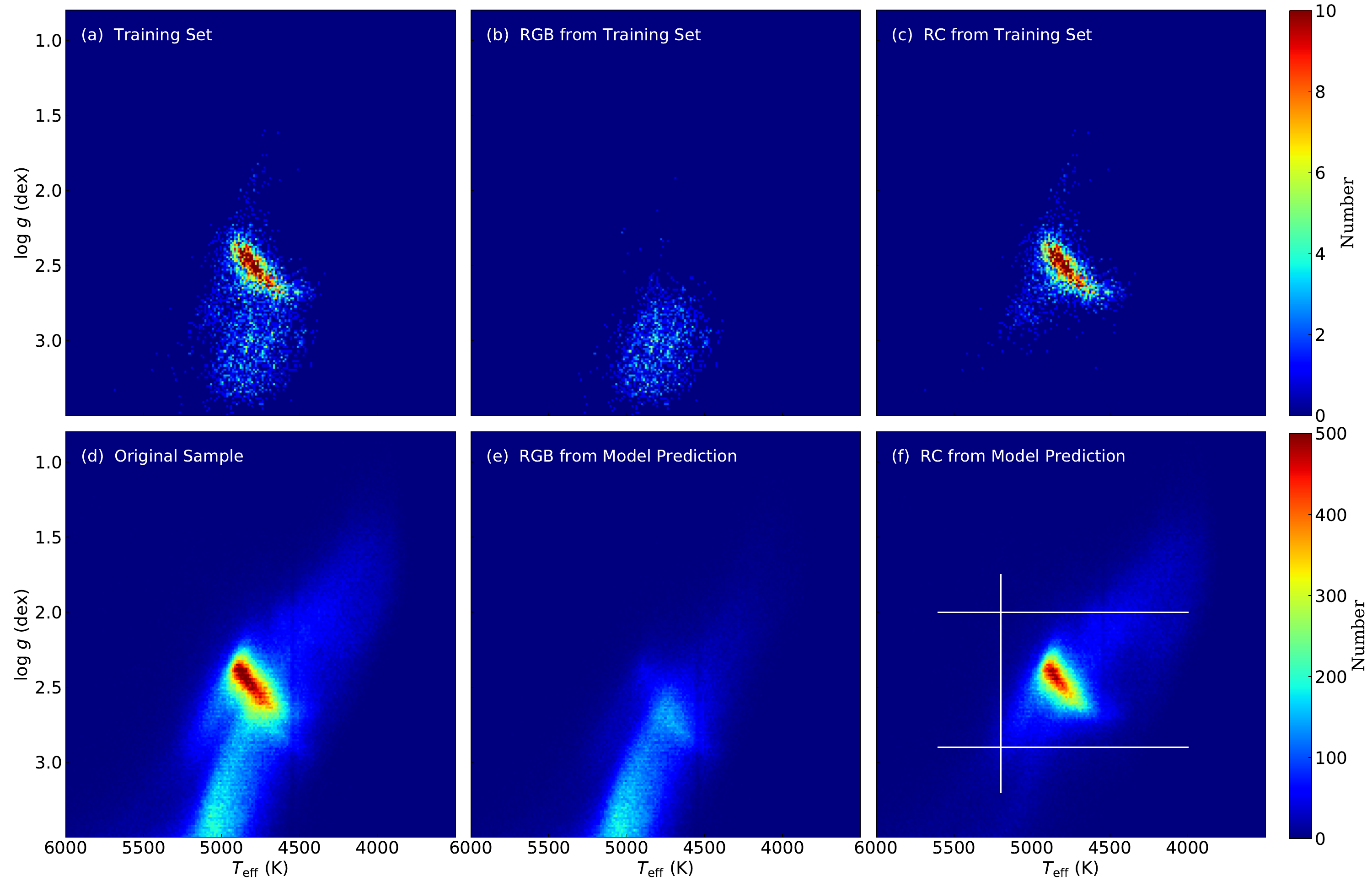}
    \caption{Distribution of training sample and classification results on original sample on $T \rm _{eff}$–log $g$ plane, binned by 12 K by 0.0135 dex and color-coded by number density. panel a): the distribution of training sample. panel b): the distribution of RGB stars from training sample. panel c): the distribution of RC stars from training sample. panel d): the distribution of original sample selected from LAMOST DR7. panel e): the distribution of RGB stars predicted by the two-class classification model. panel f): the distribution of RC stars predicted by the two-class classification model. The horizontal and vertical white lines give cuts for removing low reliability RC stars with log $g$ < 2.1 dex, log $g$ > 2.9 dex and $T \rm _{eff}$ > 5200K, respectively.}
    \label{fig:fig3}
\end{figure*}

\begin{figure}
    \includegraphics[width=\columnwidth]{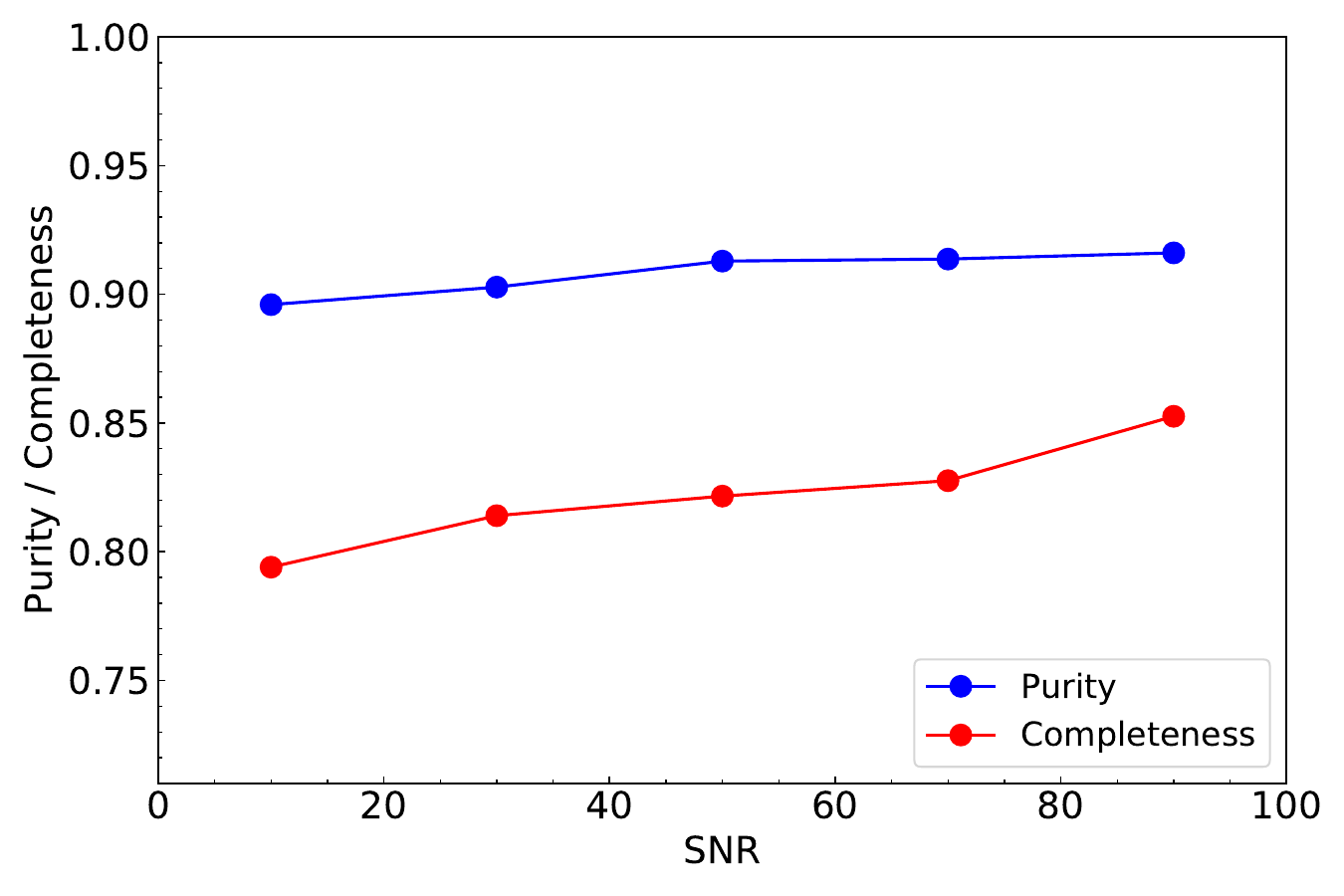}
    \caption{Purity (blue dots) and completeness (red dots) of the PRC sample with different SNR, as deduced using the asteroseismic sample with evolutionary stages classified by  \citet{elsworth2017}.}
    \label{fig:purAndCom}
\end{figure}

First, we trained the three-class classification model. According to the performance of the model in the test set, we find that the recognition performance of the model for SRC is poor. Although SRC can be accurately separated from RGB, it is not significantly different from PRC. So we put SRC and PRC as a group to train the two-class classification model, and then use other methods to eliminate SRC. Since the problem to be solved is not a simple linear problem, we choose gradient boosting tree as the structure of the model (i.e. booster: gbtree) instead of a linear function. The depth of each tree is no more than 5 (maxdepth = 5), and the L2 regularization parameter is set to 6 (lambda = 6), which helps to reduce the complexity of the model and prevent over fitting. After 300 rounds of training, the final accuracy in the test set is 96.04\%, the recall rate is 97.07\%, and the AUC is 95.53\%. The confusion matrix is shown in Fig. \ref{fig:matrix} showing that our model has high reliability for classifying most samples in the test data set.

We select the stars widely covering the RC region with log $0.9<g<3.5$ dex, $-0.6< [Fe/H] <0.4$ dex, and $3600<T \rm _{eff}<6000$ from the LAMOST DR7 catalog. Finally, we obtain 819,657 spectra (original sample) with SNRs higher than 10 as shown in panel (d) in Fig.\ref{fig:fig3}.

We classified the spectra in the original sample with the two-class classification model, and identified 377 565 RGB spectra and 442 092 RC spectra. Fig. \ref{fig:fig3} shows the distribution of training sample and classification results on the original sample on $T \rm _{eff}$–log $g$ diagram, respectively. It shows that most of the stars in the high density region of the RC stars are correctly predicted as RC. At the same time, due to the lack of RGB samples in the low temperature and low surface gravity part of the training set, the classified RC samples still have some pollution in this region. 

In order to further improve the purity of the RC sample, we only retain 80\% of the stars closest to the center, and remove the part with too sparse data. As shown in panel (f) in Fig. \ref{fig:fig3}, Then, the condition of  $T \rm _{eff}$ > 5200K, log $g$ < 2.1 dex and log $g$ > 2.9 dex are applied to remove some stars, which is used to further select RC samples with higher reliability\citep{wan2015red}. Compared with secondary RC stars, only PRC stars have a fixed magnitude because they are mainly composed of relatively old RC stars. So if we want to use RC stars to track the evolution of the Galactic disk, SRC stars should be removed  from the sample. \cite{wan2015red} combined with PERSEC stellar evolution model, and fitted the absolute magnitude of RC stars to a function of log $g$, [Fe/H], and $T \rm _{eff}$. Using the method provided by \cite{wan2015red}, we calculated the absolute magnitude of RC stars in K$_s$ band. According to the property that PRC has fixed intrinsic magnitude, we eliminated the SRC part with a linear relationship between magnitude and log $g$ in $M_K$-log $g$ diagram, and finally got 175,809 PRC spectra.

To verify the purity and completeness of the selected PRC sample, we used the catalog of the evolution stages of Kepler stars based on the asteroseismic analysis from \cite{elsworth2017} as the verification sample. The sample contains  6637 stars. After deleting the stars repeated with the training sample, we cross match them with LAMOST DR7 catalog and our PRC sample at SNR > 10. We obtained 1316 spectra (607 RGB, 597 PRC, 112 SRC) and 529 spectra (43 RGB, 474 PRC, 12 SRC), respectively. It shows that when the SNRs are greater than 10, the purity of PRC sample is 89.60\%, and the completeness is 79.40\%. The results for the purity and completeness when using higher SNR spectra are presented in Fig. \ref{fig:purAndCom}. The purity is over 90\% for SNRs higher than 30. And the completeness increases to over 85\% for SNRs higher than 90.

\subsection{Important features}
\label{sec:features}

For the current two-class classification model, we calculate the contribution of each feature to the prediction of the model. Here we choose to measure the importance of each feature by calculating the average SHAP value. After summarizing, six important wavelength ranges are obtained as shown in Table ~\ref{tab:table2}.

\begin{table}
	\centering
	\caption{The important wavelength ranges of classification model.}
	\label{tab:table2}
	\begin{threeparttable}
	\begin{tabular}{ccccc} 
	    \hline
		  & start & end & SHAP score & spectal name\\		
		\hline
		1 & 5248 & 5252 & 1.72 & Fe5270\tnote{a}\\
		\hline
		2 & 5167 & 5191 & 1.28 & MgH \& Mg I$b$\tnote{a}\\
	    \hline
		3 & 4956 & 4960 & 1.18 & Fe I 4957.6\tnote{b}\\
		\hline
		4 & 4205 & 4209 & 1.16 & Fe I 4207.1\tnote{b}\\
		\hline
		5 & 5205 & 5213 & 0.57 & Cr I 5208.4\tnote{b}\\
        \hline
		6 & 4173 & 4178 & 0.36 & CN\tnote{a}\\
		\hline
	\end{tabular}
	\begin{tablenotes}
	\item[a] denotes the line index from \citep{liu2015spectral}.
	\item[b] denotes the line index from the ispec theory line index table.
	\end{tablenotes}
	\end{threeparttable}
\end{table}

\begin{figure}
    \includegraphics[width=\columnwidth]{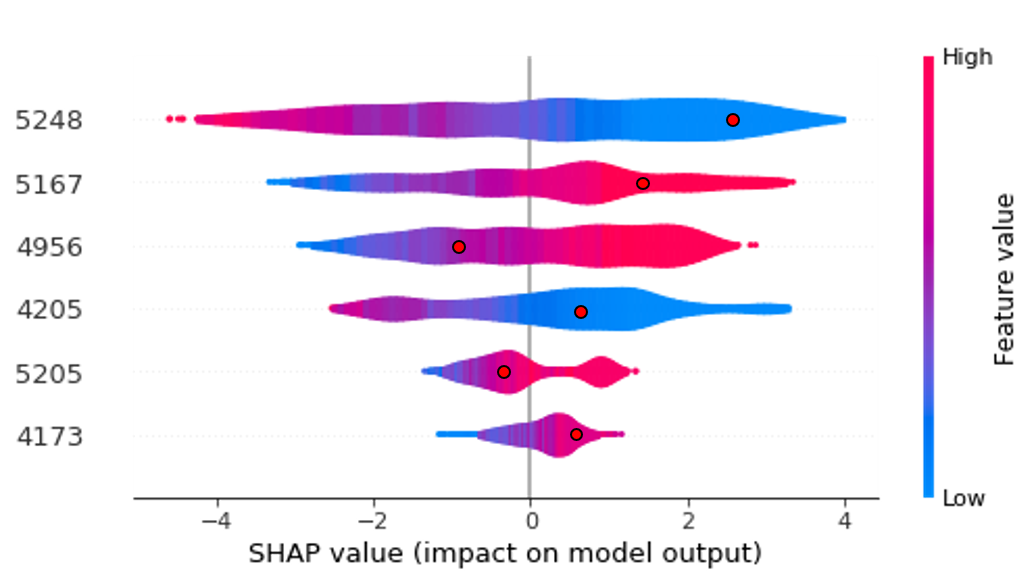}
    \caption{The contribution ranking of different features to the model derived from SHAP method. The red color represents the larger value, and blue color represents the smaller value. An example is shown in figure as a red point.}
    \label{fig:fig4}
\end{figure}

The spectral flux values of these six wavelength ranges are summed up respectively and used as new features to train a new classification model. With the same data and parameter settings as the previous classification model, the accuracy of the new model is stable at about 97\%. The corresponding SHAP scores of each feature in the new model are shown in Table ~\ref{tab:table2}, Fig. \ref{fig:fig4} shows the contribution of different features to the new model. Six features are arranged vertically, color represents the size of the feature value, red represents the larger value, and blue represents the smaller value. The horizontal axis is the calculated SHAP value of each sample at each feature. The larger the absolute value of SHAP, the greater the influence of this feature on the judgment of the model. In this model, the larger the positive shape value is, the more the model thinks that the current sample is an RC star. The larger the negative SHAP value is, the more the model tends to judge it as RGB. An example of RC spectrum is shown in Fig. \ref{fig:fig4} as a red point. It can be seen that the SHAP value of the spectrum on four features is greater than 0, and the final summation is also greater than 0. The interpretation model determines that it is an RC spectrum, which is consistent with the result given by the two-class classification model.

\begin{figure*}
    \includegraphics[width=1.6\columnwidth]{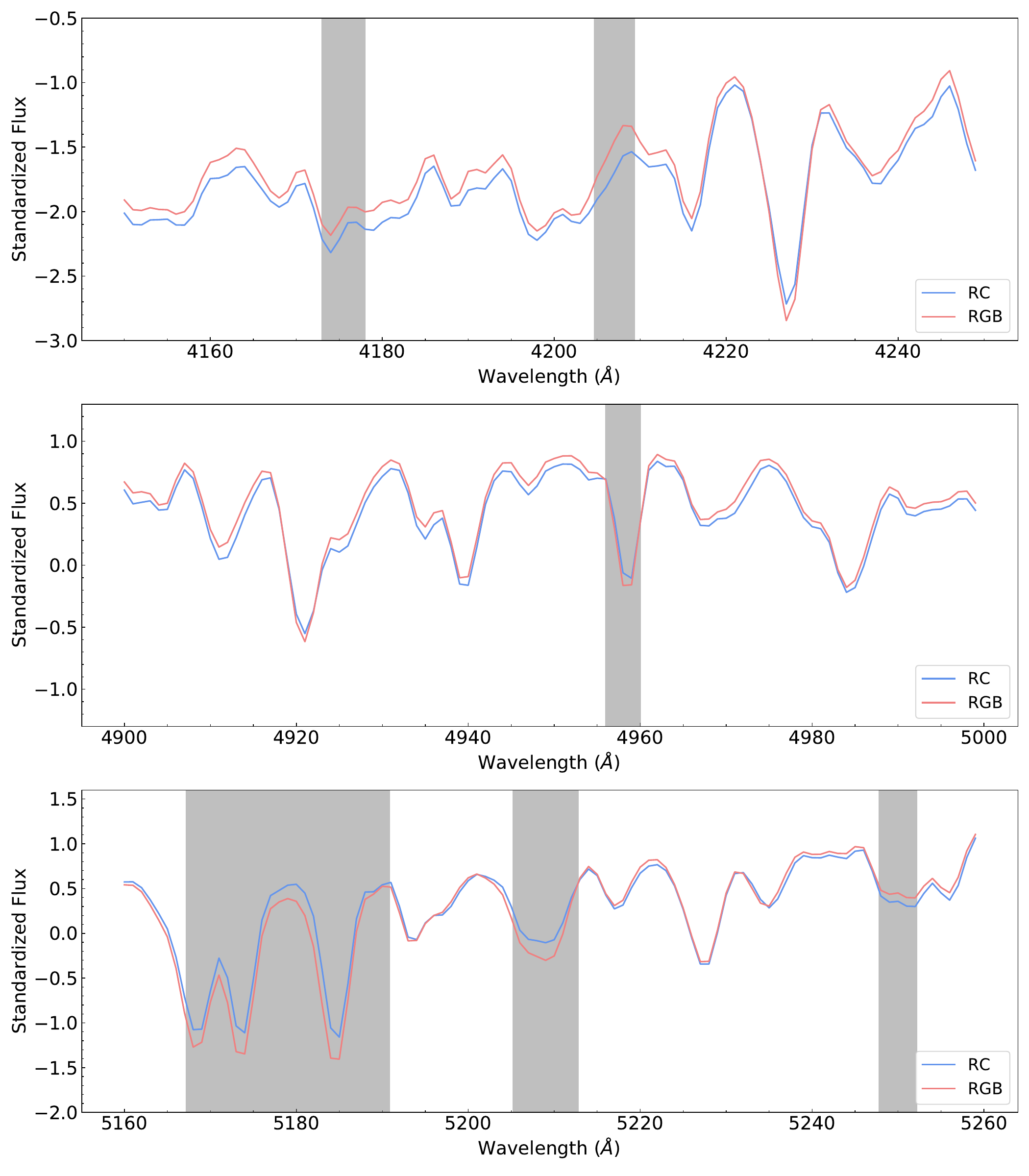}
    \caption{The average spectrum of RC and RGB stars from training set. The grey lines are the important feature bands derived from SHAP method.}
    \label{fig:feature}
\end{figure*}

\begin{figure}
    \includegraphics[width=\columnwidth]{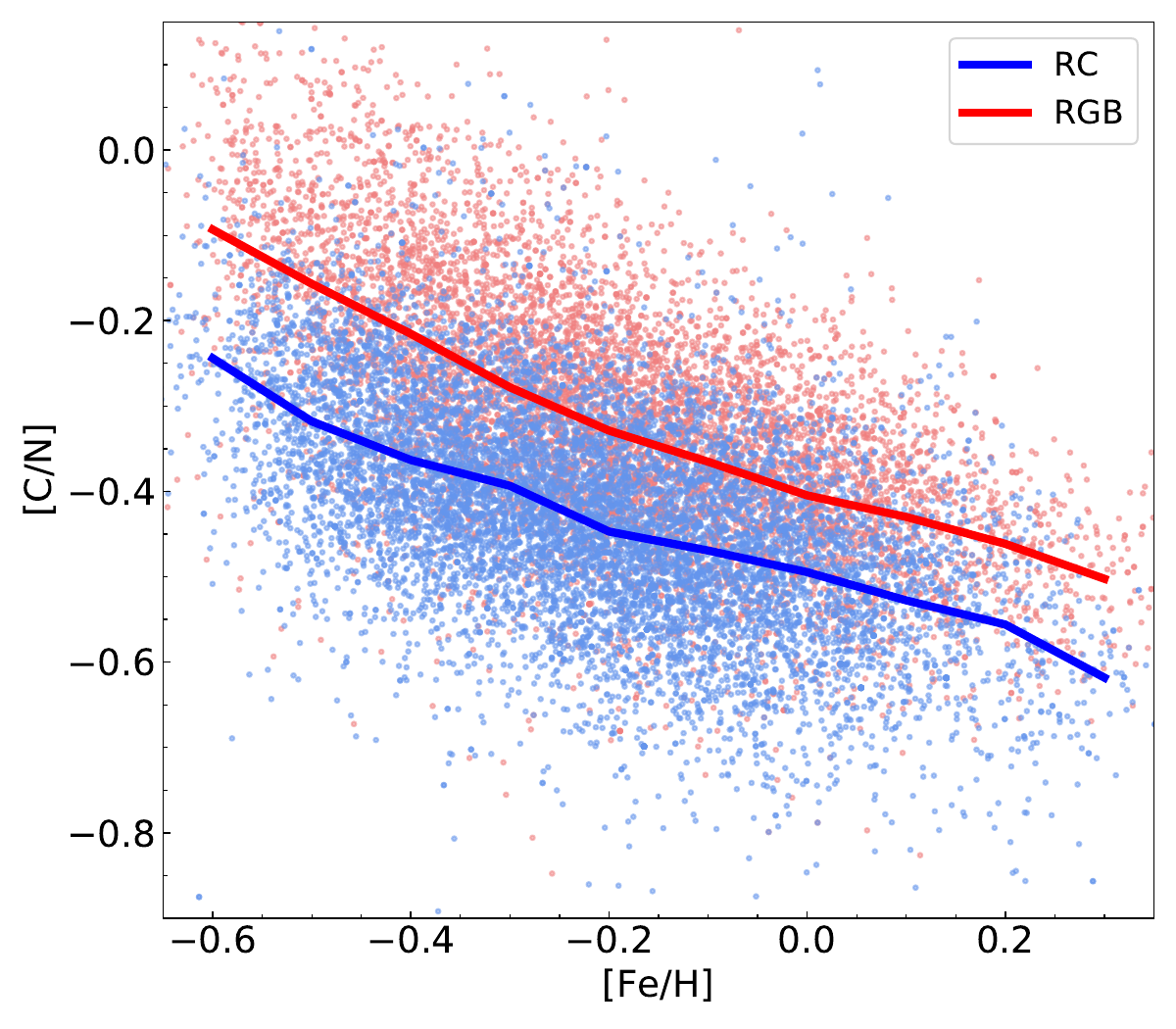}
    \caption{The [C/N] as a function of [Fe/H] for RGB (red points) and RC (blue points). The solid red and blue lines represent the mean value of the RGB and RC in this space, respectively.}
    \label{fig:feh_cn}
\end{figure}

The last column of Table ~\ref{tab:table2} is the specific spectral lines we have identified from low and medium-resolution line index tables. These important feature bands are also covered by gray rectangles on both RC and RGB average spectrum in Fig. \ref{fig:feature}. The features evaluated by our model are usually very consistent with empirical features. Three of the important spectral lines shown in Table ~\ref{tab:table2} are Fe line index, which means [Fe/H] plays an important role in the XGBoost model. At the same time, $T \rm _{eff}$-log $g$ determines the position difference of RC and RGB stars on the HR diagram. For example, \cite{bovy2014} used $T \rm _{eff}$-log $g$ diagram to divide RC and RGB, so the $T \rm _{eff}$-related and log $g$-sensitive line index dominates the classification. The Mg I triplet ($\lambda\lambda$5167, 72, 83) increases with the decrease of $T \rm _{eff}$, which is clearly shown in the low resolution spectrum, so the MgH bands become significant criteria around K5\citep{2009Stellar}. For the cool star-like RC and RGB, Mg I$b$'s Van der Waals damping broadening line wing depends on log $g$, which means there exists a correlation between the equivalent width of Mg I$b$ and log $g$. The abundance of chromium is generally related to the abundance of iron, so the ratio of the Cr I triplet to its nearby Fe I can precisely classify K-stars. \citep{2009Stellar}. For the stars cooler than 5000K, $T \rm _{eff}$ depends on iron's ionization equilibrium(Fe II / Fe I). It shows that although high-precision log $g$ and $T \rm _{eff}$ can not be obtained from LAMOST low-resolution spectrum, these differences can also be reflected in the spectrum, and star classification can be very effective by feature extraction. The C and N abundances change continuously when RGB stars evolve to RC stars, and they are also reflected in the LAMOST spectrum. Fig. \ref{fig:feh_cn} shows the [C/N] for RGB (red points) and RC (blue points) stars, which is a function of [Fe/H] and shows as solid lines. The [C/N] ratios are taken from \cite{xiang2019abundance}. The difference of the [C/N] between RGB and RC stars is about 0.2 dex which is in common with the conclusion of \cite{hawkins2018}. Since [C/N] is age-sensitive, we can expect that age (and mass) can be determined from spectral features such as CN. It has been found that age-sensitive [$\alpha$/Fe] ratios are slightly different between RC and RGB (e.g. low giant branch) stars as shown in Fig.~1 of \cite{zhao2021low} based on \cite{xiang2019abundance}.

\begin{figure}
    \includegraphics[width=\columnwidth]{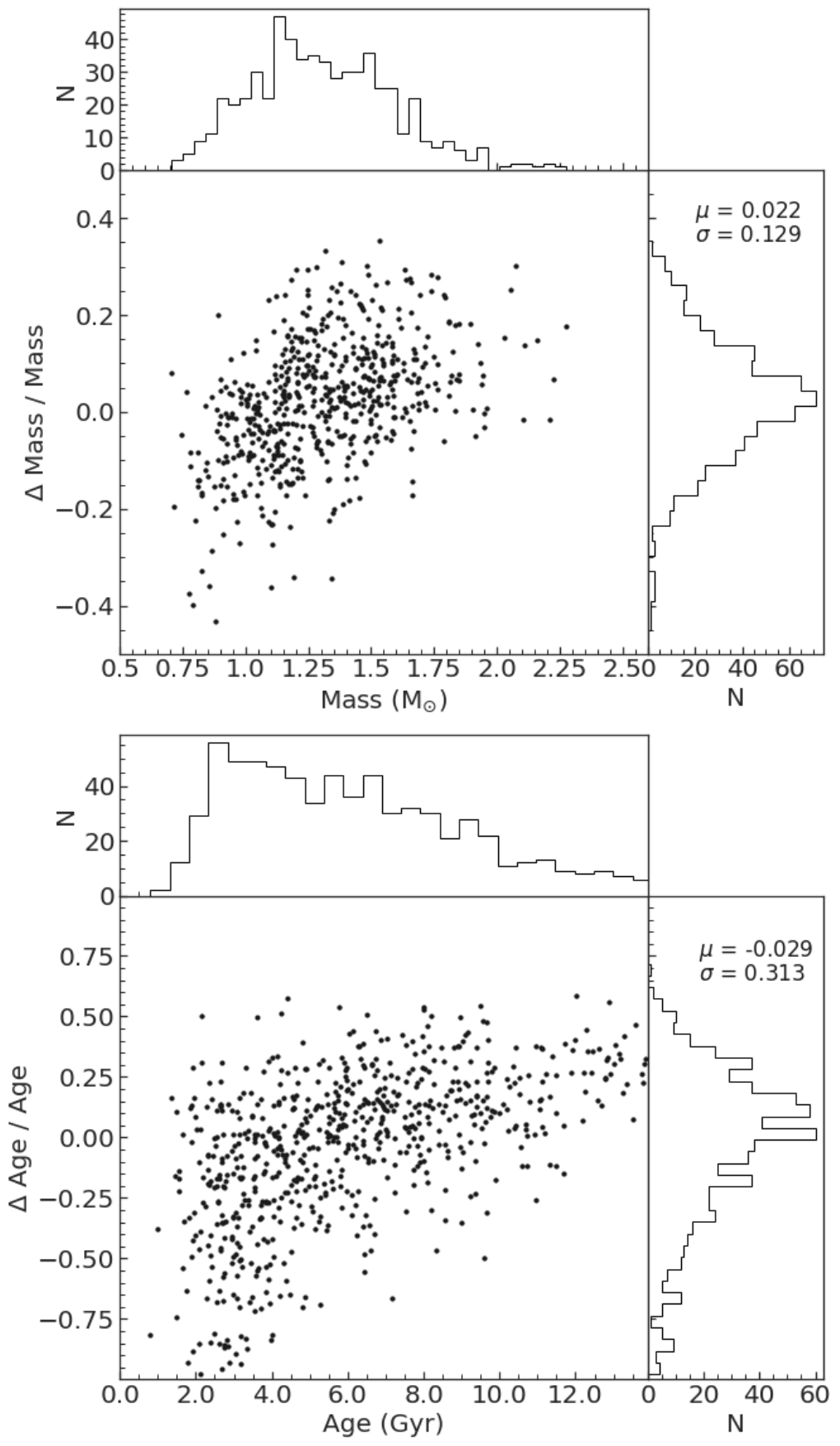}
    \caption{Upper panel: distribution of mass estimates and their errors. Bottom panel: distribution of age estimates and their errors.}
    \label{fig:fig5}
\end{figure}

\begin{figure}
    \includegraphics[width=0.95\columnwidth]{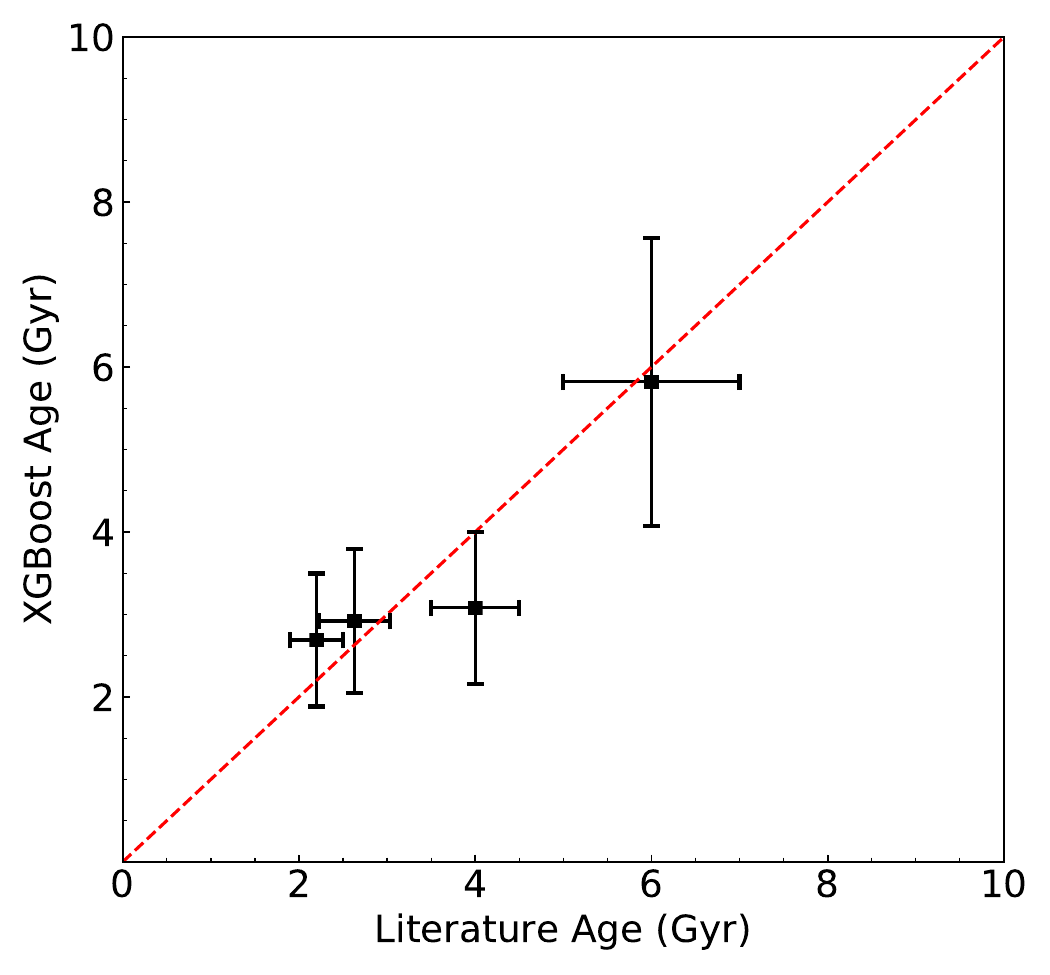}
    \caption{The comparison of XGBoost ages with the literature values for open clusters. }
    \label{fig:fig6}
\end{figure}

\section{Determination of mass and age}
\label{sec:age}

Mass and age are important information to understand the stellar population, but they can not be observed directly. In this section, we estimate the mass and age for the selected PRC stars. First, we construct a training set with precise mass and age labels. The mass and age are calculated by combining the $T \rm _{eff}$, log $g$ and [Fe/H] from LAMOST DR7 catalog, the seismological parameters provided by \cite{yu2018} and the PERSEC stellar evolution model \citep{bressan2012parsec}. We cross the two catalogs and obtain 5336 common resources with SNRs higher than 20. Based on previous studies, we used the modified relation to obtain the mass of the stellar sample by means of the stellar $T \rm _{eff}$, the asteroseismic parameters, and the modified parameters $\triangle \mu_{\odot}$ which were obtained from \cite{sharma2016stellar} is performed as

\begin{equation}
    \frac{M}{M_\odot}=(\frac{\triangle \nu}{f_{\triangle \nu}\triangle \nu_\odot})^{-4}(\frac{\nu_{\rm max}}{\nu_{\rm max, \odot}})^{3}(\frac{T \rm _{eff}}{T \rm _{eff, \odot}})^{3/2},
	\label{eq:t4}
\end{equation}
where $T \rm _{eff, \odot}=5777K$, $\nu_{max,\odot}=3090\mu \rm Hz$ and $\triangle\nu_{\odot}=135.1\mu \rm Hz$ \citep{huber2011testing}.

Based on the above estimated mass and stellar atmospheric spectral parameters ($T \rm _{eff}$, log $g$ and [Fe/H]), the ages of these PRC stars can be further deduced from stellar isochrons. We use the Bayesian method from \cite{xiang2017} and the PERSEC stellar isochrones \citep{bressan2012parsec}. The input parameters are mass M, surface gravity log $g$, effective temperature $T \rm _{eff}$ and metallicity [Fe/H]. The basic parameters of stars we used are from LAMOST DR7 catalog. We calculate the mass and age of 5336 stars and then select the star samples whose SNR is greater than 50, the mass error is less than 15\%, and age error is less than 40\% as the samples for further model training and validation. The 3079 selected stars are classified into training set and verification set according to 7:3. The sample features are LAMOST blue-arm spectrum (3900 – 5500\AA), and XGBoost algorithm is used to train the model. After 300 rounds of training, the performance of the mass and age models in the test set are shown in Fig. \ref{fig:fig5}. The RMSE of the two models are 0.169 and 1.573 respectively, and the MAE is 0.131 and 1.179 respectively.

Open clusters (OCs) are objects composed of hundreds to thousands of stars with weak gravitational connections. They are usually formed by the same huge molecular cloud. Therefore, the members of the same open cluster are very similar in age and chemical composition, which can be used to verify the age of stars. The number of very old open clusters is small in the footprint of LAMOST and they are usually dimmer. As a result, they can hardly be observed by LAMOST. We selected four open clusters in a wide age range (2–8 Gyr) visited by LAMOST for verification. Fig. \ref{fig:fig6} directly compares the ages between XGBoost gives and from literature. It can be seen from the figure that the ages of XGBoost essentially agree with previous literature. In NGC 2420, three members LAMOST observed are PRC candidates, and their mean age is determined as 2.69 Gyr, and the dispersion is 0.81 Gyr, consisting with the literature (2.2 Gyr; \citealt{demarque1994gap,twarog1999zeroing}). In NGC 6819, the mean age of six PRC candidates is 2.92 Gyr, and the dispersion is 0.87 Gyr. It is also consistent with the ages given in the literature (2.6 Gyr; \citealt{sarajedini1999wiyn,grocholski2002wiyn}). For NGC2682, we find only one PRC star, and the age is 3.08 Gyr. We use the prediction error of XGBoost model to represent its dispersion which is 0.92 Gyr. It is underestimated by about 1.00 Gyr compared with the ages given in the literature (4.0 Gyr; \citealt{sandquist2004high,an2007distances,heiter2014metallicity,stello2016k2}). While for Berkeley 32, the mean age of the two PRC is 5.82 Gyr with a dispersion of 1.74 Gyr, which is consistent with the ages given in the literature (6.0 Gyr; \citealt{salaris2004age,d2006old,tosi2007old}). 

For the mass model and the age model, we calculate the contribution of each feature to the model prediction. We still used SHAP to measure the importance of each feature. After summarizing all the features, the most important wavelength ranges are selected which are shown in Table ~\ref{tab:table4} and Table ~\ref{tab:table5}. For PRC stars, there is a correlation between their mass and age, because more massive stars will evolve more quickly and be younger when at the red clump evolutionary stage than less massive stars. \citet {2000Kong} apply the method of principal component analysis to a sample of simple stellar populations and find some important age sensitive spectral indices such as G4300, C$_2$4686, MgI$b$, MgH, Fe4383, Fe5335, and Fe5270. Besides the well-known age-sensitive index H$\beta$, we also find some new age sensitive, such as Fe4198.3 and H$\delta$.


\begin{figure}
    \centering
	\begin{minipage}{\linewidth}
		\centering
		\includegraphics[width=\linewidth]{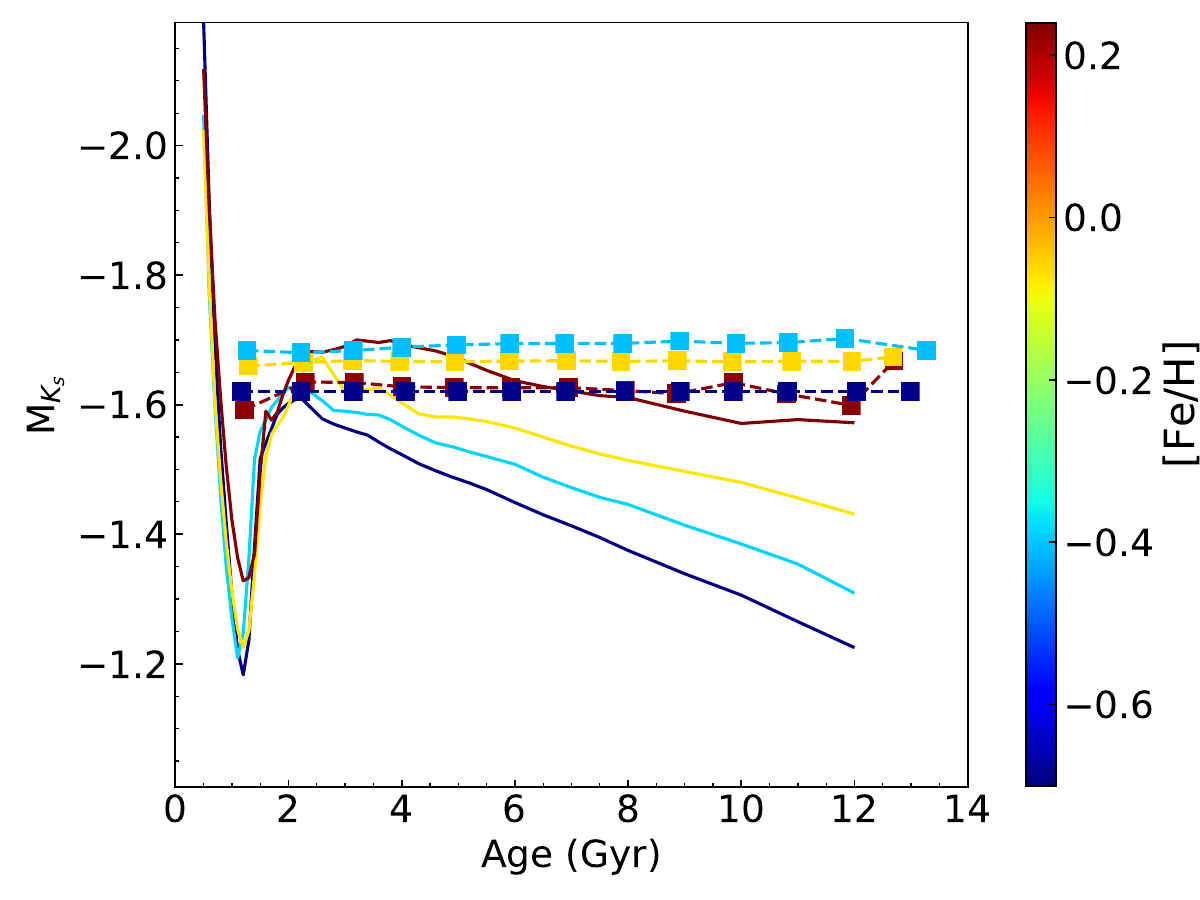}
	\end{minipage}
	
	\begin{minipage}{\linewidth}
		\centering
		\includegraphics[width=\linewidth]{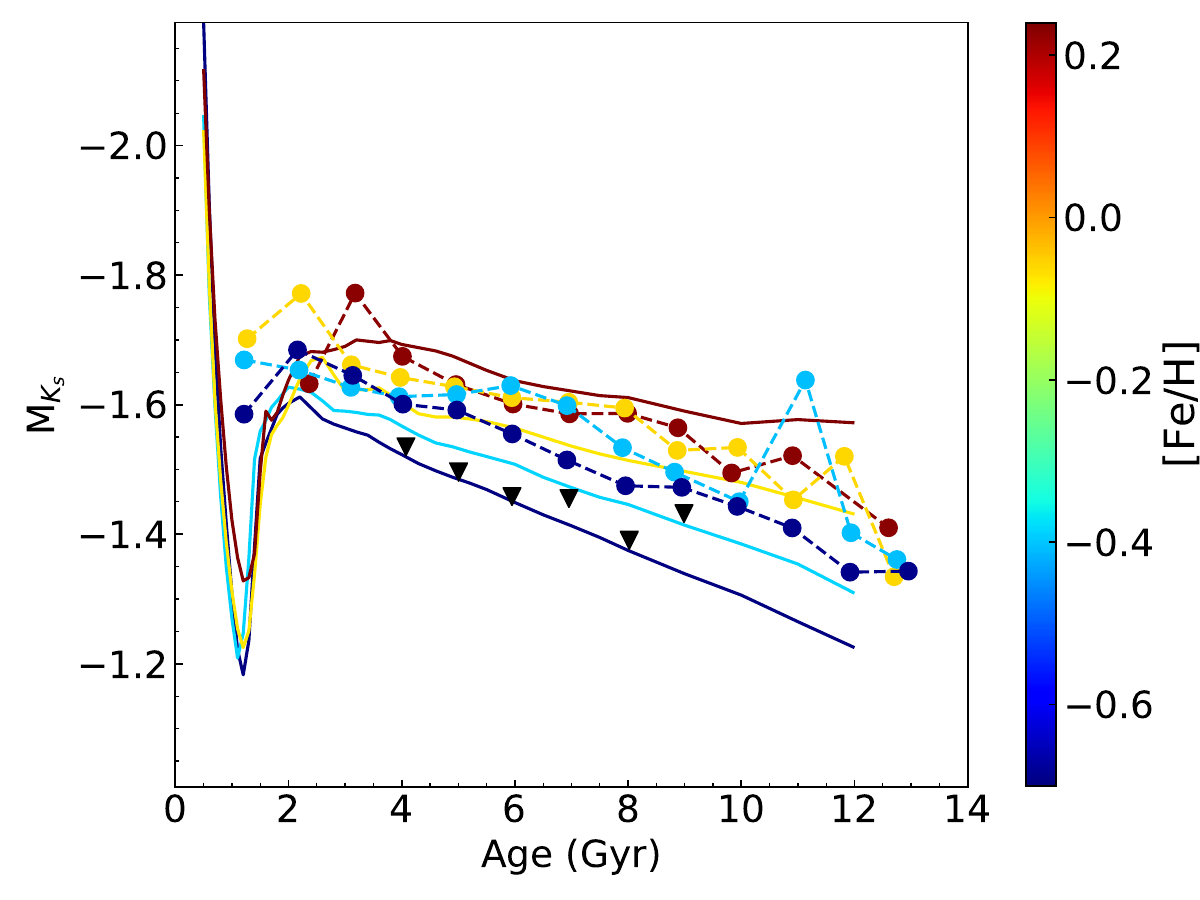}
	\end{minipage}
	\caption{Upper panel: The model predicted absolute magnitudes in K$_S$-band compared with our absolute magnitudes. Squares represent data we obtained by the method from \citet{wan2015red}. Bottom panel: The model predicted absolute magnitudes in K$_S$-band compared with Gaia EDR3. Dots represent data we obtained from the Gaia EDR3. The inverted triangle in black are the stars with [Fe/H] from -0.6 to -0.57 dex. Solid lines present the K$_S$-band mean absolute magnitude of red clump stars as a function of age and metallicity provided by \citet{salaris2002population} derived from the isochrones of \citet{girardi2000evolutionary}. The color bar represents the mean [Fe/H] of stars.}
	\label{fig:fig7}
\end{figure}

\begin{figure}
    \includegraphics[width=0.95\columnwidth]{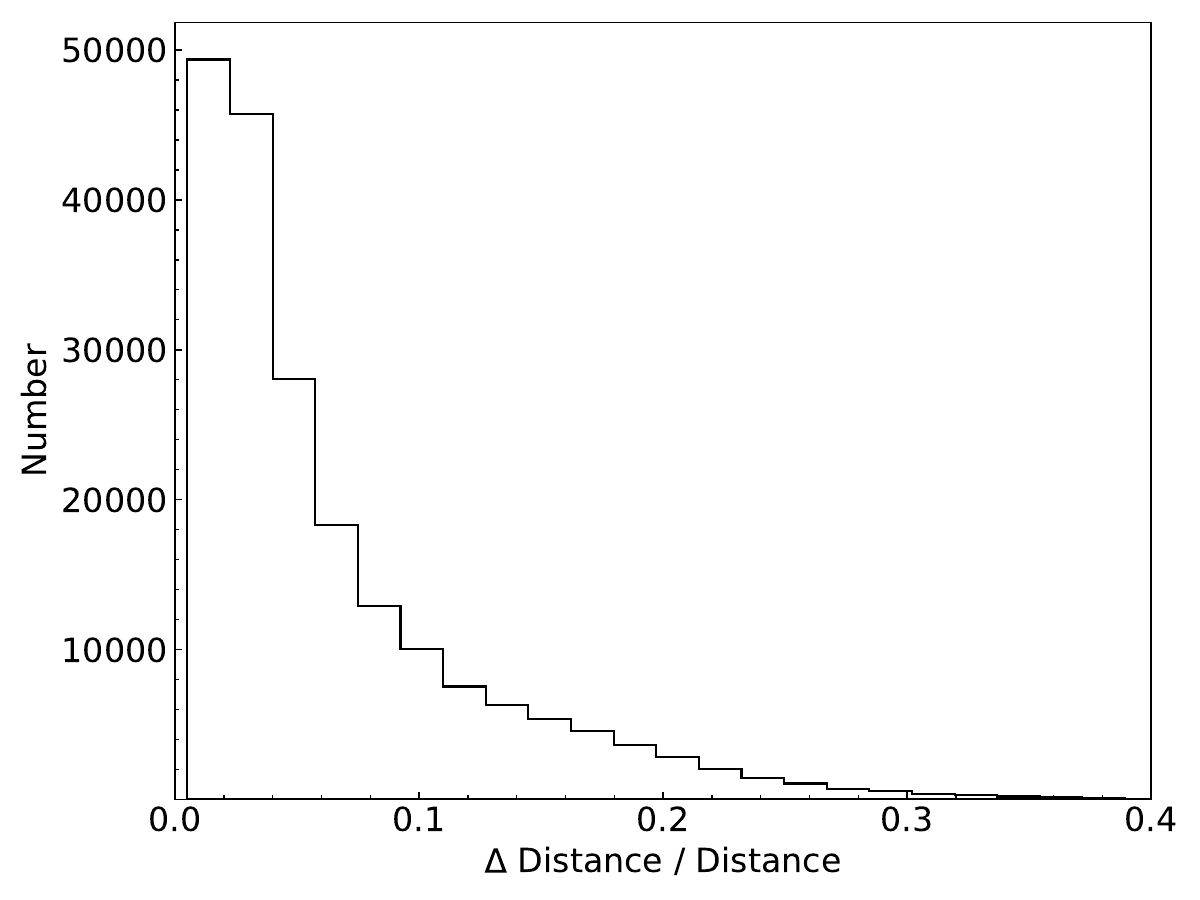}
    \caption{Intrinsic error distribution of distance derived using $K_s$ band absolute magnitude.}
    \label{fig:uncerdis}
\end{figure}


\begin{table}
	\centering
	\caption{The important wavelength ranges of mass model.}
	\label{tab:table4}
	\begin{threeparttable}
	\begin{tabular}{ccccc} 
		\hline
		  & start & end & SHAP score & spectal name\\
		\hline
		1 & 5165 & 5166 & 0.132 & MgH \& Mg I$b$\tnote{a} \\
		\hline
		2 & 4554 & 4555 & 0.114 & Fe4531\tnote{a}\\
		\hline
		3 & 4178 & 4181 & 0.102 & CN\tnote{a}\\
		\hline
		4 & 4312 & 4313 & 0.102 & G4300\tnote{a}\\
        \hline
		5 & 5129 & 5131 & 0.068 & Fe I 5131.5 \& Ti I 5129.2\tnote{b}\\
		\hline
		6 & 3970 & 3976 & 0.042 & Fe I 3971.3\tnote{b}\\
		\hline
	\end{tabular}
	\begin{tablenotes}
	\item[a] denotes the line index from \citep{liu2015spectral}.
	\item[b] denotes the line index from the ispec theory line index table.
	\end{tablenotes}
	\end{threeparttable}
\end{table}

\begin{table}
	\centering
	\caption{The important wavelength ranges of age model.}
	\label{tab:table5}
	\begin{threeparttable}
	\begin{tabular}{ccccc}
		\hline
		  & start & end & SHAP score & spectal name\\		
		\hline
		1 & 5269 & 5273 & 1.270 & Fe5270\tnote{a}\\
		\hline
		2 & 5164 & 5166 & 1.253 & Mg I$b$\tnote{a}\\
		\hline
		3 & 5128 & 5131 & 0.994 & MgH\tnote{a}\\
        \hline
		4 & 5325 & 5327 & 0.656 & Fe5335\tnote{a}\\
		\hline
		5 & 4195 & 4198 & 0.629 & Fe I 4198.3\tnote{b}\\
		\hline
		6 & 4859 & 4863 & 0.512 & H$\beta$\tnote{a}\\
		\hline
		7 & 4301 & 4305 & 0.438 & G4300\tnote{a}\\
		\hline
		8 & 4395 & 4396 & 0.392 & Fe4383\tnote{a}\\
		\hline
		9 & 4312 & 4312 & 0.382 & G4300\tnote{a}\\
		\hline
		10 & 4078 & 4080 & 0.325 & H$\delta$\tnote{a}\\
		\hline
		11 & 4373 & 4375 & 0.294 & Fe4383\tnote{a}\\
		\hline
	\end{tabular}
	\begin{tablenotes}
	\item[a] denotes the line index from \citep{liu2015spectral}.
	\item[b] denotes the line index from the ispec theory line index table.
	\end{tablenotes}
	\end{threeparttable}
\end{table}

\begin{figure*}
    \includegraphics[width=1.7\columnwidth]{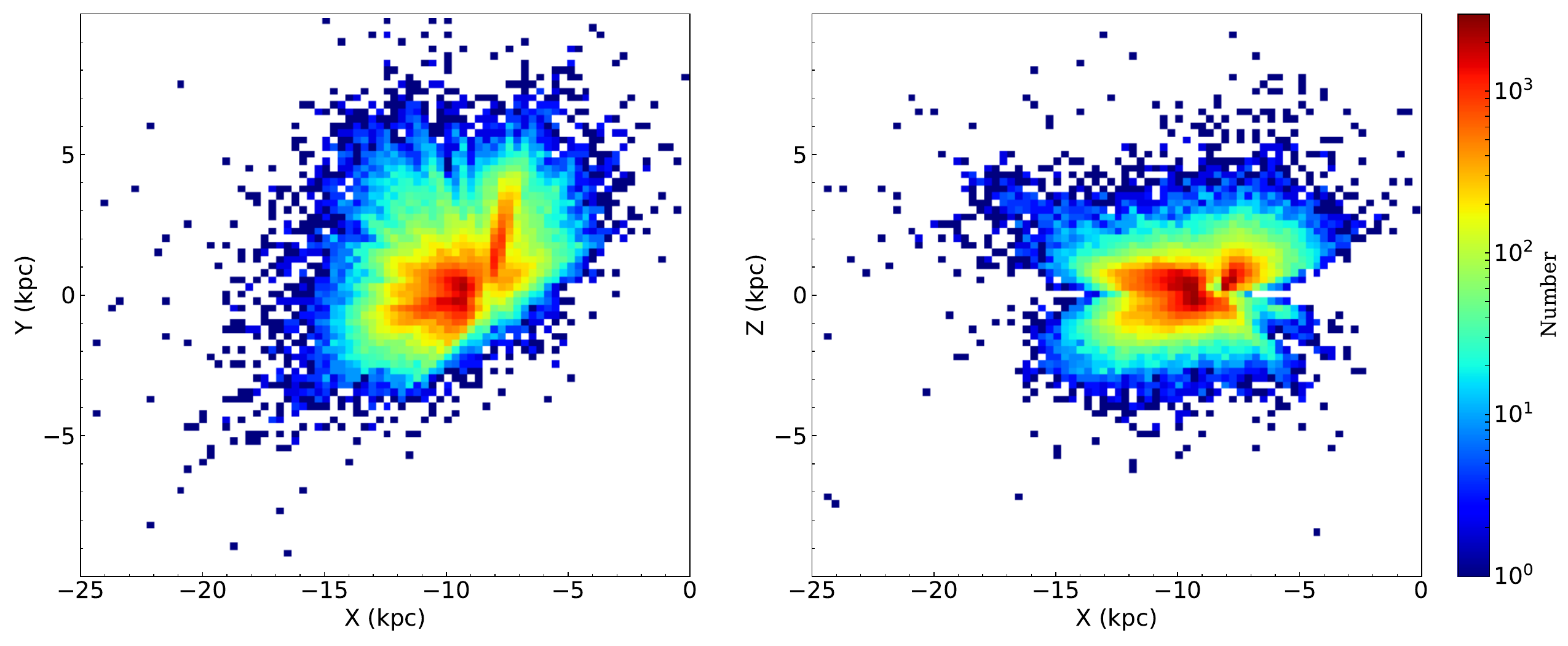}
    \caption{Number density distribution of our primary RC sample stars in the X-Y (left panel) and X-Z (right panel) planes. The stars are binned by 0.25 × 0.25 kpc$^2$. The densities are shown on a logarithmic scale.}
    \label{fig:density-xyz}
\end{figure*}

\begin{figure*}
    \includegraphics[width=2\columnwidth]{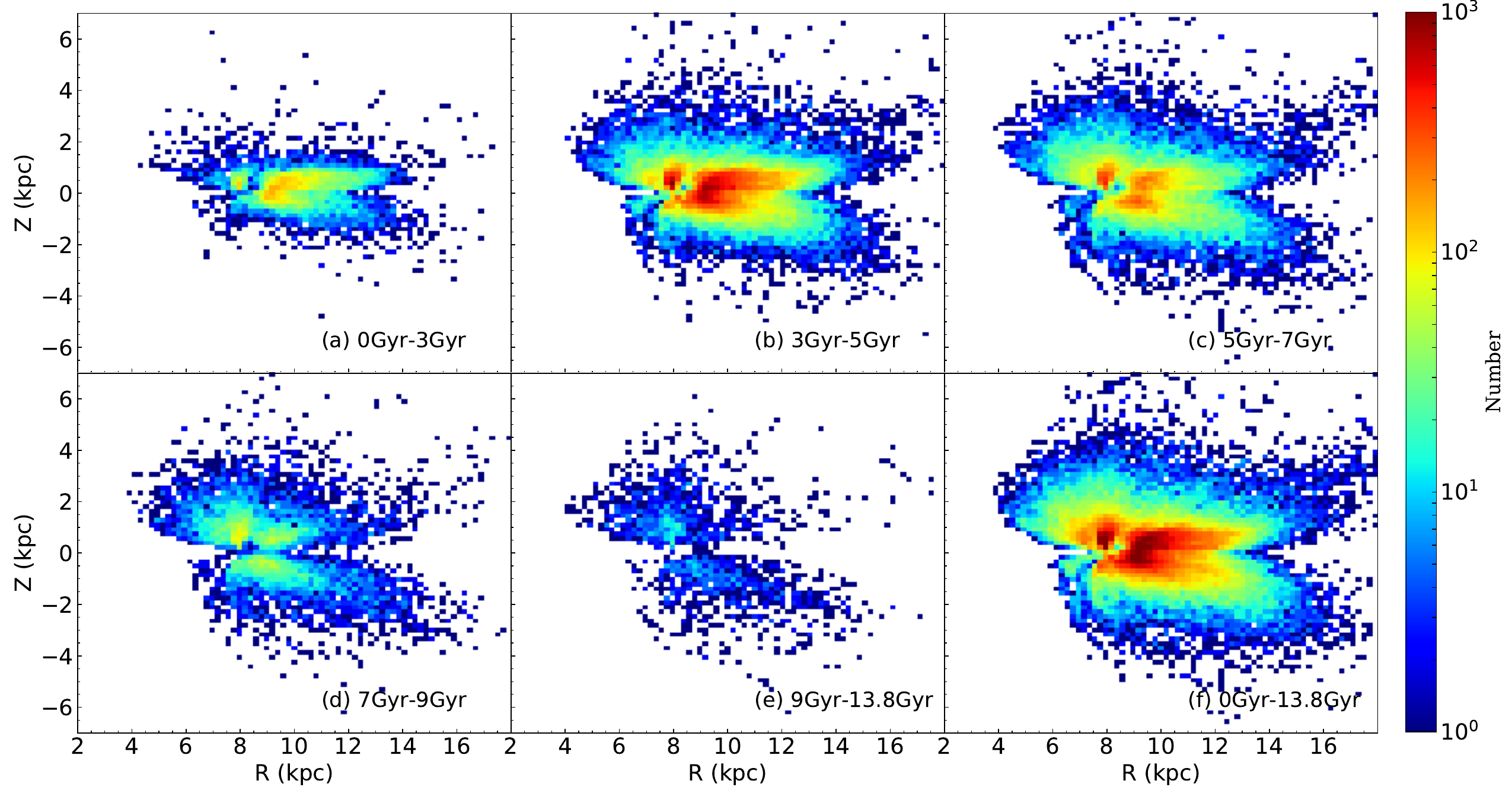}
    \caption{Distributions of our RC sample stars in the R–Z plane, binned by 0.25 × 0.25 kpc$^2$ and color-coded by number density with different age intervals. panel a): the distribution with age from 0 Gyr to 3 Gyr. panel b): the distribution with age from 3 Gyr to 5 Gyr. panel c): the distribution with age from 5 Gyr to 7 Gyr. panel d): the distribution with age from 7 Gyr to 9 Gyr. panel e): the distribution with age from 9 Gyr to 13.8 Gyr. panel f): the distribution with age from 0 Gyr to 13.8 Gyr. The densities are shown on a logarithmic scale.}
    \label{fig:agebin-r-z}
\end{figure*}

\section{Determination of distance}
\label{sec:distance}

Accurate measurement of distance is one of the most basic and difficult goals of astronomy. Using RC as a distance indicator began with the work of \cite{cannon1970red}. Almost every work on stellar evolution has well confirmed that RC has intrinsic absolute magnitude and can be used as a good standard candle. Compared with other bands, the absolute magnitude in the near-infrared $K_s$ band is least affected by the population effect and interstellar reddening. However, even in the near-infrared band, ignoring the population effects may still introduce 5-10\% systematic error \citep{salaris2002population,girardi2016}. The distance modulus $\mu_0 = 5 log d - 5$ is performed as
\begin{equation}
    \mu_0 = m^{RC}_\lambda - M^{RC}_\lambda - A_\lambda + \triangle M^{RC}_\lambda,
	\label{eq:t5}
\end{equation}
where $\lambda$ represents the corresponding band. We use $K_s$ band to calculate the distance. $m^{RC}_\lambda$ represents the apparent magnitude of the star from 2MASS \citep{2003yCat.2246....0C,skrutskie2006two}, $A_\lambda$ represents the interstellar extinction in $\lambda$ band, and we use the three-dimensional extinction map of the galaxy provided by \cite{green20193d}. $\triangle M^{RC}_\lambda$ is a ``population correction" to correct the unavoidable presence of the population effects, a dependence of stellar absolute magnitude on its age.

To derive distances of the PRC stars, an accurate calibration of absolute magnitudes is required. We calculated a magnitude for each star according to its atmospheric parameters using the function from \citet{wan2015red} as we introduced in section \ref{sec:classification}. So the magnitudes we obtained from section \ref{sec:classification} are equivalent to the ``$M^{RC}_\lambda + \triangle M^{RC}_\lambda$" in formula \ref{eq:t5}. However, the dependence of stellar absolute magnitude on its age is still not obvious as shown in the upper panel of Fig. \ref{fig:fig7}. The dependency of absolute magnitude of RC on metallicity and mass (or age) is previously reported in \cite{2001ApJ...551L..85Z} and \cite{2017ApJ...840...77C}.To calibrate our $K_s$ absolute magnitude, we constructed a sample of common stars of our PRC sample and the Gaia EDR3 \citep{bailer2021estimating}. We used some cuts to improve reliability of the sample such as, the stars have LAMOST spectra of SNRs higher than 60, the photometric errors in the $K_s$ band smaller than 0.03 mag, the relative parallax uncertainties smaller than 10 percent and the parallax uncertainties smaller than 0.05 mas. We obtained 65,666 common stars and their $K_s$ absolute magnitudes derived from the Gaia EDR3 distances \citep{bailer2021estimating}. The Gaia EDR3 magnitudes show obvious dependence on ages as shown in the bottom panel of Fig. \ref{fig:fig7}. Because the [Fe/H] of our sample are higher than -0.6 dex, the observed data appear higher position than the theoretical line especially the dark blue line. The inverted triangle in black is the average $K_s$ absolute magnitude of stars with [Fe/H] from -0.6 to -0.57 dex, which is close to the the dark blue line. We used second-order polynomial to fit the $K_s$ absolute magnitude in the upper and bottom panel of Fig. \ref{fig:fig7} as a function of metallicity and age, respectively. Then use the discrepancy between our magnitude and the Gaia EDR3 magnitude to calibrate our magnitude. For the distances derived from $M_{K_S}$ calibrated by using the Gaia EDR3 data, the resulting distances are biased less than 10\%.

We use the error propagation formula to derive the error of estimated distance from the error of atmospheric parameters and apparent magnitude. As shown in the Fig. \ref{fig:uncerdis}, the uncertainty of the estimated distance is mostly within 10\%, and a few (about 3\%) have an uncertainty greater than 20\%. The average uncertainty is about 6\%.

\section{The primary RC sample}
\label{sec:sample}

Based on the data-driven method, we obtain 175,809 PRC stars from the LAMOST DR7 with mass, age distance, and metallicity. Combined with RA and DEC provided by the LAMOST catalog, we calculated the axes of a Galactocentric (X, Y, Z) and Galactocentric distance (R) of these PRC stars. Using this PRC sample, we explored the spatial distribution of stars in the galaxy, and the relationship between the star mass, age, and metallicity and their spatial position. Fig. \ref{fig:density-xyz} shows the density distributions of PRC stars in the Galactic X-Y and X-Z planes. Here, the X, Y, and Z are computed assuming $R_{\odot}$ = 8.299 kpc and $Z_{\odot}$ = 27 pc. The primary RC sample covers a large volume of the Galactic disk, X from -20 to 0 kpc, Y from -5 to 5 kpc, and Z from -5 to 5 kpc. Therefore, the spatial distribution, stellar population distribution, and chemical kinematic characteristics of PRC stars can reflect the characteristics of the Galactic disk to a certain extent.

Fig. \ref{fig:agebin-r-z} shows the density distribution of stars with different age intervals in the R-Z plane. For young stars between 0 Gyr and 3 Gyr, most of them are concentrated in the range of - 1.5 kpcs to 1.5 kpcs in the Z direction. Young stars are mainly born near the Galactic disk, and there is no obvious density change in the R direction. For stars between 3 Gyr and 5 Gyr, there are more stars in the region within 2 kpcs in the Z direction, and the number density of stars in the higher region also increases. For stars from 5 Gyr to 7 Gyr, the stars in the Z direction expand to higher regions, and the density of stars within 2 kpc decreases.The number density of stars near the Galactic disk decreases with age. The figure shows that young stars are mainly distributed near the Galactic disk, while old stars are more evenly distributed both in the vicinity of the Galactic disk and halo.

\begin{figure}
    \includegraphics[width=\columnwidth]{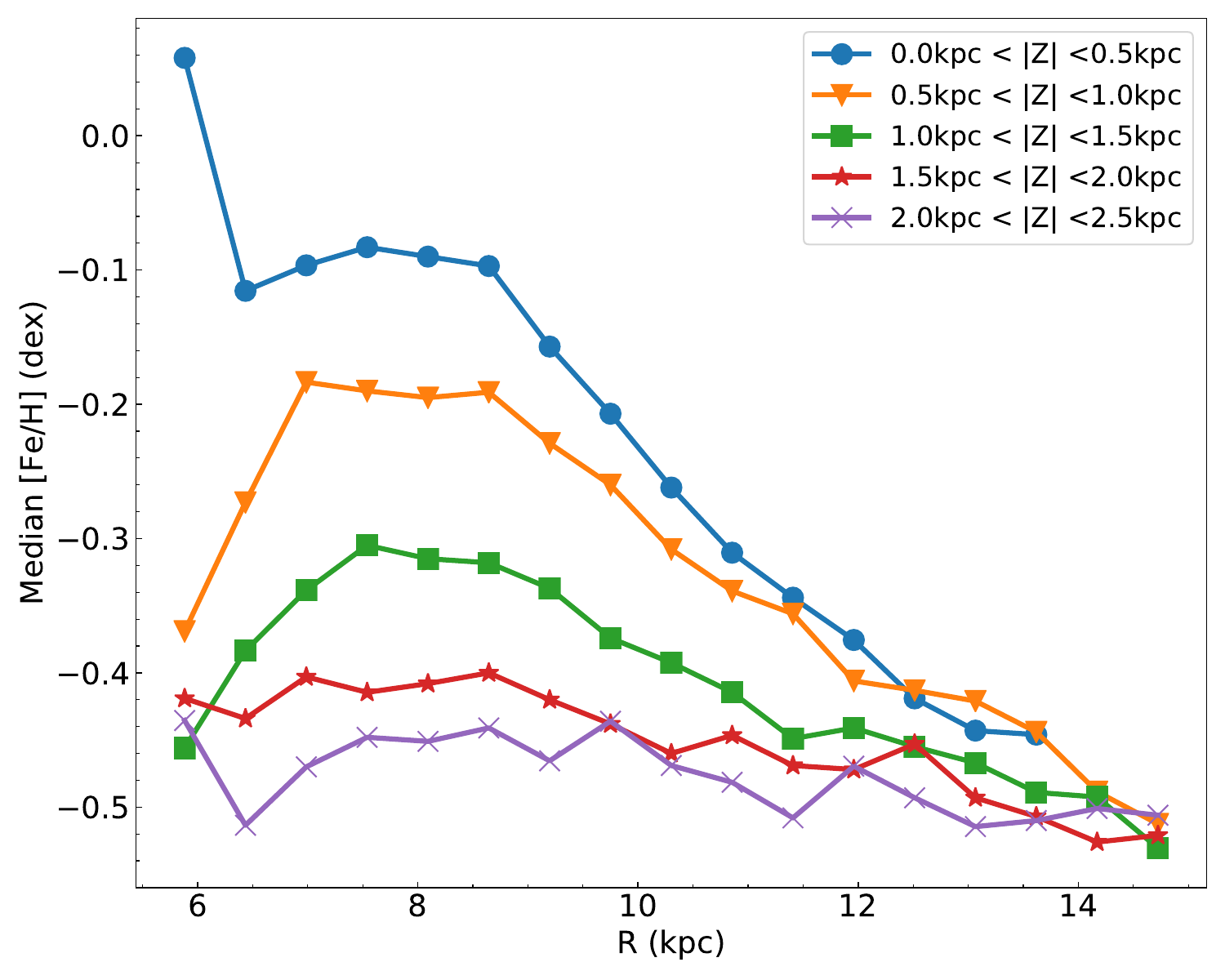} 
    \caption{The R-[Fe/H] relation in different bins of |Z| coded by colors. |Z| = 0 kpc: the blue lines;  |Z| = 0.5 kpc: the orange lines;  |Z| = 1.0 kpc: the green lines, |Z| = 1.5 kpc; the red lines; and |Z| = 2.0 kpc: the purple lines.}
    \label{fig:zbin-feh-r}
\end{figure}

\begin{figure*}
    \includegraphics[width=2\columnwidth]{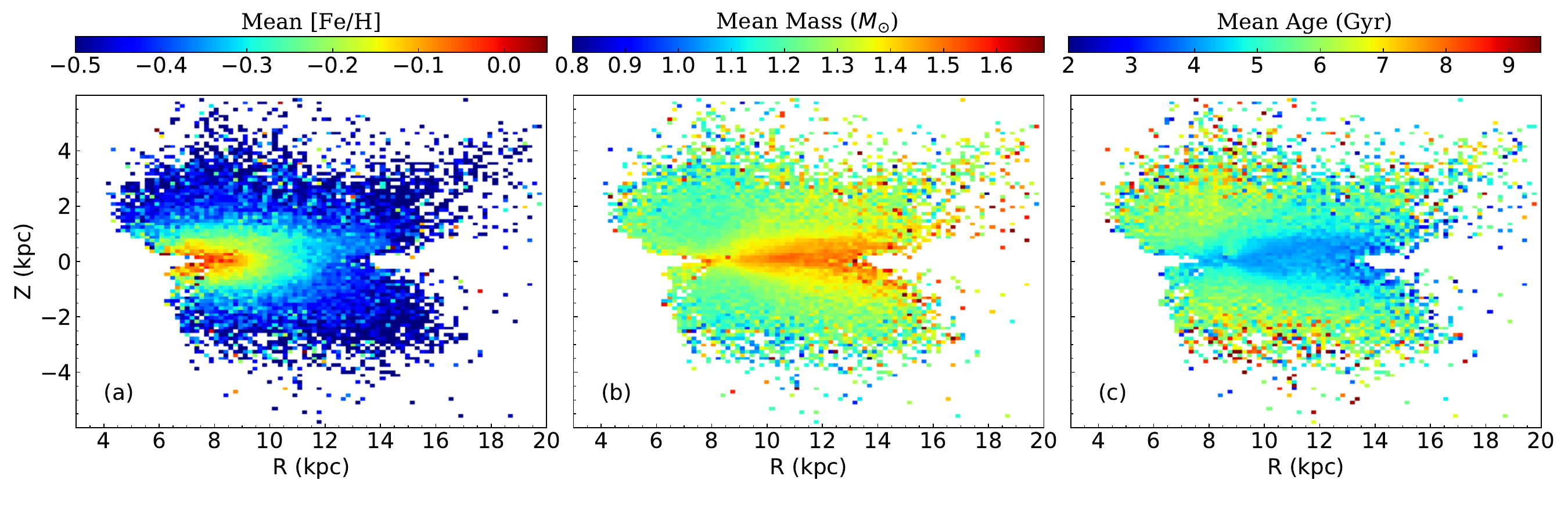} 
    \caption{Distributions of our RC sample stars in the R-Z plane, binned by 0.20 dex by 0.15 dex and color-coded by mean [Fe/H] (Panel a), mean mass (Panel b) and mean age (Panel c).}
    \label{fig:distribution}
\end{figure*}

Fig. \ref{fig:zbin-feh-r} shows the metal abundance of stars at different |Z| changes with R. From 6 to 15 kpcs, as R increases, in each |Z| interval, the metal abundance shows a downward trend. The metal abundance dropped from -0.1 to -0.5, indicating that there is a gradient of metal abundance from the center of the Milky Way to the more peripheral regions. From 0 to 2.5 kpcs, with the increase of |Z|, the metal abundance at different Rs shows an overall downward trend, indicating that there is a gradient of metal abundance from the plane of the Milky Way to higher regions. Panel (a) of Fig. \ref{fig:distribution} shows the mean values of [Fe/H] at different positions across the R-Z plane of the Galactic disk. It shows an obvious chemical thick disk, and most stars in the Galactic halo are metal-poor stars. 

Panel (b) and (c) of Fig. \ref{fig:distribution} show the mean values of mass and age at different positions across the R-Z plane of the Galactic disk. Young Galactic disk is clearly detected in both distributions, where stars are younger and more massive. Due to the mass loss during the evolution of PRC stars, the mass and age in the two distributions are inversely correlated. Mass values show a negative gradient and age values show a positive gradient in the vertical direction. As the R increases, the younger populations extended to higher heights of the Galactic plane.

\section{Discussion and Summary}
\label{sec:discussion}

\begin{figure*}
    \includegraphics[width=1.9\columnwidth]{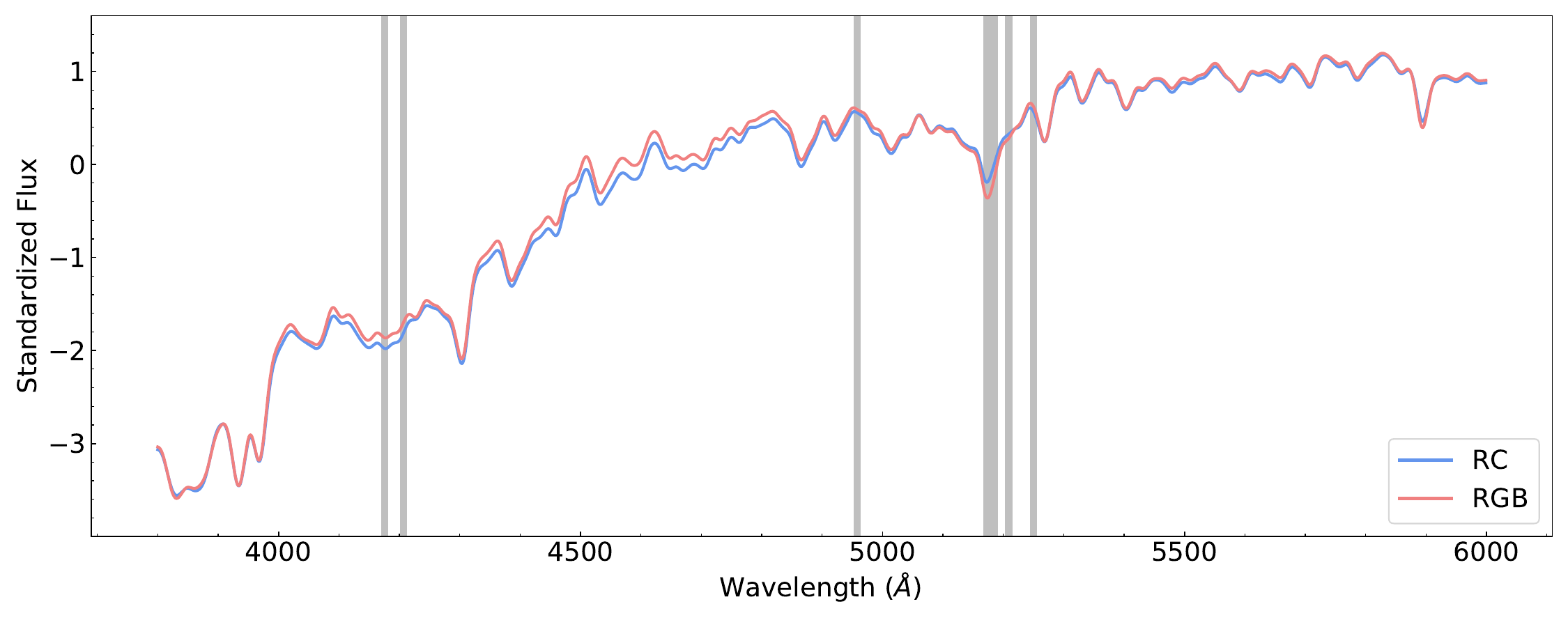}
    \caption{The average spectrum of RC and RGB stars from training set with 250 resolution. The grey lines are the important feature bands derived from SHAP method.}
    \label{fig:low-feature}
\end{figure*}

To further verify the visibility of spectral features in low-resolution spectrum, we reduced the resolution of the spectrum in the training set from 1800 to 250, and the new average spectrum after the reduced resolution is shown in Fig. \ref{fig:low-feature}. Then we used the six features of the new spectrum to train another two-class classification model, and the final accuracy in the test set reached 93\%, which means the important spectral features still can be recognized by machine learning model even with a resolution as low as 250. Taking into account that the spectral features will become fuzzier in the lower resolution spectrum, a model with higher accuracy can be obtained by appropriately expanding the wavelength range when calculating the features. This technology may be applied to data mining of the next generation low-resolution sky surveys, such as CSST (China Space Station Optical Survey Telescope; \cite{gong2019cosmology}).

In this work, nearly 185,000 primary RC stars have been successfully identified from the spectra with SNR higher than 10 of LAMOST DR7 , based on the XGBoost model trained through thousands of RC and RGB stars having Kepler high-quality asteroseismology data. The sample has been verified with their positions in the metallicity-dependent $K_S$ absolute magnitude–surface gravity diagrams. Cross-validation with Kepler targets shows that the purity and completeness of the primary RC sample are higher than 90 percent and 80 percent, respectively. The catalog can be found at \url{http://paperdata.china-vo.org/hexujiang/prc/catalog_of_primary_red_clump.csv}. We provided the LAMOST spectrum id, the position, the atmospheric parameters, the apparent magnitude in J, H and K bands and their errors of the PRC sample. Likewise, we derived and provided the $K_s$ absolute magnitude, the mass, the age, the distance and their errors of the PRC sample which called ``mk", ``mass", ``age" and ``dist" in the catalog.

The SHAP feature attribution method is used to explain the XGBoost two-class classification model, which shows the difference in the spectrum of RGB  and RC stars is mainly reflected in six spectral ranges which correspond to Fe5270, MgH $\&$ Mg I$b$, Fe4957, Fe4207, Cr5208, and CN. 

The ages and masses of primary RC stars are determined from the LAMOST spectra with the XGBoost method, trained with stars with accurate asteroseismic parameters from the LAMOST-Kepler fields. The uncertainties of the mass and age are 13 and 31 percent, respectively. After the verification of the SHAP method, the age-related spectral features are consistent with the age-sensitive elements that have been recorded in previous literature.

We use 2MASS absolute magnitude of  $K_{S}$ band to estimate distances of the PRC sample with an uncertainty of about 6 percent. Cross-matching with the distance measured by \citep{bailer2018estimating}using parallaxes published in the second Gaia data release, the deviation of the the two results are 19 pc and the dispersion of the residual is 752 pc.

We conclude that the XGBoost presented in this study provides an an efficient way to derive a vast and high purity sample of RC, especially for low-resolution spectra because of their low precision of effective temperature and surface gravity measurement which hard to separate RC and RGB stars using traditional methods. Besides, it can also be used to estimate mass, age, and distance for PRC stars. More importantly, it has the capability to extract important spectral features related to stellar type, mass, and age. The primary RC sample we derived using XGBoost covers a significant volume of the Galactic disk, allowing us to map the structure, kinematics, and evolution of the Galactic disk in future works. 

\section*{ACKNOWLEDGEMENTS}

This work is supported by the National Key R\&D Program of China No. 2019YFA0405502, the National Natural Science Foundation of China (Grant Nos. U1931209, 11988101,11890694, and 12011530055), and the science research grants from the China Manned Space Project with NO.CMS-CSST-2021-B05. This research has made use of LAMOST data. The full name of LAMOST is the Large Sky Area Multi-Object Fiber Spectroscopic Telescope or Guoshoujing Telescope. It is a National Major Scientific Project built by the Chinese Academy of Sciences. Funding for the project has been provided by the National Development and Reform Commission. LAMOST is operated and managed by the National Astronomical Observatories, Chinese Academy of Sciences.



\section*{Data Availability Statements}
The data underlying this article are available in the China-VO Paper Data Repository at \url{http://paperdata.china-vo.org} and can be accessed with this link \url{http://paperdata.china-vo.org/hexujiang/prc/catalog_of_primary_red_clump.csv}.

\bibliographystyle{mnras}
\bibliography{article.bib} 

\bsp	
\label{lastpage}
\end{document}